\RequirePackage{fix-cm}
\documentclass[onecolumn]{svjour3}
\smartqed 
\usepackage[utf8]{inputenc}          %
\usepackage{graphicx}
\usepackage{xcolor}

\usepackage{booktabs}
\usepackage{siunitx}
\usepackage{subcaption}
\usepackage{graphicx}
\usepackage{amsmath}
\usepackage[export]{adjustbox}
\usepackage{pythonhighlight}
\usepackage{caption}
\usepackage{multirow}

\usepackage[round]{natbib}
\usepackage[hyphens]{url}	

\usepackage{xurl}

\usepackage{float}

\usepackage{setspace}
\makeatletter
\let\cl@chapter\undefined
\makeatletter
\usepackage[capitalise]{cleveref}
\usepackage[margin=1in,includefoot,footskip=30pt]{geometry}

\usepackage{lineno}

\begin{document}

\title{Multiphase lattice Boltzmann modeling of cyclic water retention behavior in unsaturated sand based on X-ray Computed Tomography}
\titlerunning{Multiphase LBM simulation of hysteresis in unsaturated sands}    

\author{Qiuyu Wang\textsuperscript{*} \and Marius Milatz \and Reihaneh Hosseini \and Krishna Kumar}

\institute{
      Q. Wang, R. Hosseini \& K. Kumar \at
The University of Texas at Austin \\
Civil, Architectural and Environmental Engineering \\
Cockrell School of Engineering \\
301 E. Dean Keeton St. C2100, Austin, Texas 78712-2100, USA\\
\email{wangqiuyu@utexas.edu}
\and
M. Milatz  \at
Hamburg University of Technology (TUHH) \\
Institute of Geotechnical Engineering and Construction Management \\
       Harburger Schloßstraße 36, 21079 Hamburg, Germany \\
}


\newcommand{\ie}{\textit{i.\,e.},~}
\newcommand{\eg}{\textit{e.\,g.},~}
\newcommand{\etc}{\textit{etc.}}
\date{}
\maketitle

\begin{abstract}
The water retention curve (WRC) defines the relationship between matric suction and saturation and is a key function for determining the hydro-mechanical behavior of unsaturated soils. We investigate possible microscopic origins of the water retention behavior of granular soils using both Computed Tomography (CT) experiment and multiphase lattice Boltzmann Method (LBM). We conduct a CT experiment on Hamburg sand to obtain its WRC and then run LBM simulations based on the CT grain skeleton. The multiphase LBM simulations capture the hysteresis and pore-scale behaviors of WRC observed in the CT experiment. Using LBM, we observe that the spatial distribution and morphology of gas clusters varies between drainage and imbibition paths and is the underlying source of the hysteresis. During drainage, gas clusters congregate at the grain surface; the local suction increases when gas clusters enter through small pore openings and decreases when gas clusters enter through large pore openings. Whereas, during imbibition, gas clusters disperse in the liquid; the local suction decreases uniformly. Large pores empty first during drainage and small pores fill first during imbibition. The pore-based WRC shows that an increase in pore size causes a decrease in suction during drainage and imbibition, and an increase in hysteresis.

\end{abstract}

\section{Introduction}
\label{sec:intro}


The Water Retention Curve (WRC) of an unsaturated soil describes the relationship between its degree of saturation (${S_{r}}$) and suction ($s$). WRC is useful in determining the hydraulic conductivity, shear strength, compressibility, and swelling potential of unsaturated soils~\citep{millington1961permeability,mualem1976new,vanapalli1996model,fredlund1993soil}.

\begin{figure}[htbp]
\begin{center}
 \includegraphics[width=0.6\columnwidth]{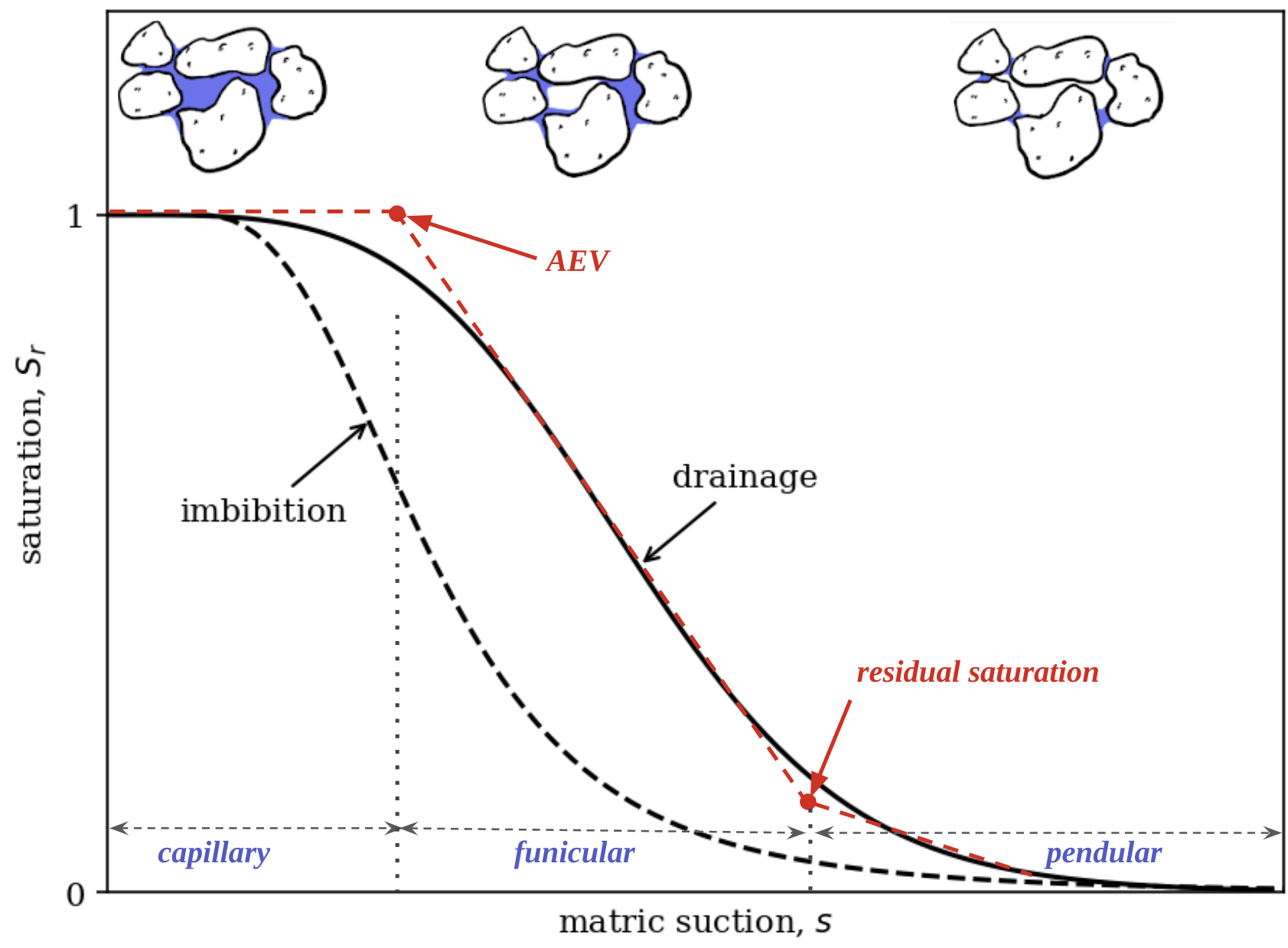}
\end{center}
\caption{Schematic of a typical WRC with a drainage and imbibition path and the corresponding pore-scale fluid distributions for the pendular, funicular, and capillary regimes.}
\label{Fig:typical_WRC}
\end{figure}

The WRC shows a hysteretic behavior, showing higher suction during drainage than imbibition for the same degree of saturation. \Cref{Fig:typical_WRC} shows a characteristic S-shaped profile of WRC with the corresponding fluid distribution in the pore space at different stages along the primary drainage and main imbibition paths. As the saturation in the pore space changes, the liquid distribution exhibits three distinct regimes: capillary, funicular, and pendular. In the \emph{capillary regime}, the soil is in a nearly fully saturated state with a few entrapped gas bubbles. A reduction in the degree of saturation leads to a significant increase in capillary pressure until the suction reaches the air-entry value (AEV)~\citep{fredlund1993soil} (see~\Cref{Fig:typical_WRC}). AEV is the suction at which the largest pore in the soil starts to desaturate during drainage. As the saturation decreases further, the liquid distribution enters the \emph{funicular regime}, where individual gas bubbles connect to form gas clusters while liquid begins to group in partially connected clusters. Beyond the funicular regime, reduction in saturation leads to the formation of binary capillary bridges between grain contacts - termed as the \emph{pendular regime} \citep{fredlund2012unsaturated,rumpf1962strength,schubert1982kapillaritat,urso1999pendular}. At residual saturation the liquid is discontinuously and exists only as a coating on the grains that can only be removed through evaporation, such as oven drying \citep{vanapalli1998meaning}. The capillary pressure at the residual saturation may reach one million kPa \citep{fredlund1994equations}.

At the microscopic scale, the air entrapment, ink-bottle effect, and contact angle hysteresis (\ie the difference in solid-liquid contact angle between drying and wetting) are proposed as possible reasons for hysteresis~\citep{mualem1974conceptual,poulovassilis1970hysteresis,ravikovitch2002experimental,de2008effect,likos2004hysteresis}. Air entrapment refers to a phenomenon during imbibition where gas bubbles are stuck in dead-end pores, causing lower suction~\citep{poulovassilis1970hysteresis}.~\citet{haines1930studies} attributed the hysteresis behavior to the geometry of the pore space - termed as the ``ink-bottle'' effect. At the same suction, wider pores with narrow neck opening can retain more liquid during drainage, leading to higher saturation than imbibition.~\citet{likos2004hysteresis} observed when the contact angle increases in an advancing meniscus during imbibition, it causes a larger radius of curvature and therefore lower suction than in a receding meniscus at the same saturation during drainage. In this study, we explore the micromechanics that contribute to the hysteretic WRC using in situ CT measurements and multiphase lattice Boltzmann Method (LBM).

The WRC of a soil is typically obtained through laboratory testing, which involves applying a capillary pressure to a soil sample in stages and measuring the degree of saturation through outflow measurements such as the hanging liquid column technique \citep{ASTM_Hanging_Water_Column}. Alternatively, liquid content or degree of saturation is controlled, and suction is measured as a response \citep{Milatz2020,Milatz_et_al2018b,Milatz_et_al2018a}. However, these laboratory experiments offer no insight on the spatial and temporal distribution of the multiphase system. In situ X-ray CT analysis offers the ability to distinguish liquid and gas at the pore scale, in conjunction with mechanical testing in a triaxial cell, uniaxial compression tests, or a flow cell-based water retention test~\citep{higo2018pore,khaddour2015multi,Milatz_et_al2021,higo2016local,kido2020morphological,Milatz_et_al2022}. In situ X-ray testing is relatively costly and time-consuming.

Many empirical models have been proposed to establish the hysteretic WRC through an analysis of pore and pore-liquid geometry. The models proposed by~\citet{fredlund2012unsaturated} and~\citet{van1980closed} involve curve-fitting which allows estimating the WRCs using the grain size distribution.~~\citet{pham2005study} applied the laboratory-measured gravimetric WRC and shrinkage curves for estimating hysteretic WRCs for sand and clay. These empirical models, however, only account for hysteresis by varying contact angles; they do not properly account for the micromechanics that contribute to the macroscopic WRCs.

In this study, we evaluate the pore-scale behavior using a multiphase numerical method. Lattice Boltzmann (LBM) is a mesoscopic multiphase method for simulating fluid flow behavior of an unsaturated soil \citep{shan1993lattice,Huang2015,kruger2017lattice}. Multiphase LBM successfully reproduces capillary effects in porous media and WRC; it also offers insights on the microscopic distribution and localized suction measurements at the pore scale~\citep{Hosseini_et_al2021,hosseini2022investigating,galindo2013lattice,scheuermann2020dynamics}. There are four main categories of multiphase LBM models: the color-gradient model; the Shan–Chen (SC) model; the free energy (FE) model; and the interface tracking model
\citep{Huang2015,gunstensen1991lattice,shan1993lattice,swift1995lattice,he1999lattice}. The SC model is the most widely used and employed in this study as it is simple, efficient, and accurately tracks the liquid-gas interface~\citep{Huang2015,kruger2017lattice,shan1993lattice}.~\citet{schaap2007comparison} showed that multiphase LBM could reproduce WRCs when gravitational, inertial, and viscous forces are negligible compared to capillary forces.\textcolor{black}{~\citet{delenne2015liquid},~\citet{richefeu2016lattice} and~\citet{galindo2016boundary} examined the cluster statistics, \ie the distribution of liquid clusters and the local grain environments to reproduce water retention curve.} However, detailed study investigating pore-scale properties such as phase distribution, interfacial areas, and pore size dependency of WRC behaviors in real sand is lacking.

In this study, we adopt a mesoscale single-component multiphase lattice Boltzmann method to simulate the macroscopic water retention curve of Hamburg sand based on 4D CT data from an in situ flow cell experiment \citep{Milatz2020,Milatz_et_al2022}. First, we successfully reproduce the WRC from the in situ CT experiment with multiphase LBM. Second, we compute microscopic properties, including liquid and gas distribution, interfacial areas, to compare the microscopic similarities and differences between CT and LBM. We also explore the pore-scale mechanisms that control the hysteresis by examining the number and size of liquid and gas clusters, morphological change in gas clusters, and liquid pressure distribution in LBM. We finally present the dependence of water retention behaviors on local pore size.

\section{Material and methods}
\label{sec:material_and_methods}

\subsection{Hamburg sand}
\label{subsec:investigated_sand}

We investigate the WRC for a medium to coarse grained Hamburg sand \citep{Milatz2020,Milatz_et_al2018b,Milatz_et_al2018a}. Hamburg sand represents a model soil widely used at Hamburg University of Technology (TUHH). 
The sand grains are mainly well rounded with smooth quartz surfaces, although angular grains with sharp edges due to fracture or mineral inclusions are also present.

\begin{figure}
  \centering
  \begin{subfigure}[b]{.7\textwidth}
    \centering
    \adjincludegraphics[width=\columnwidth]{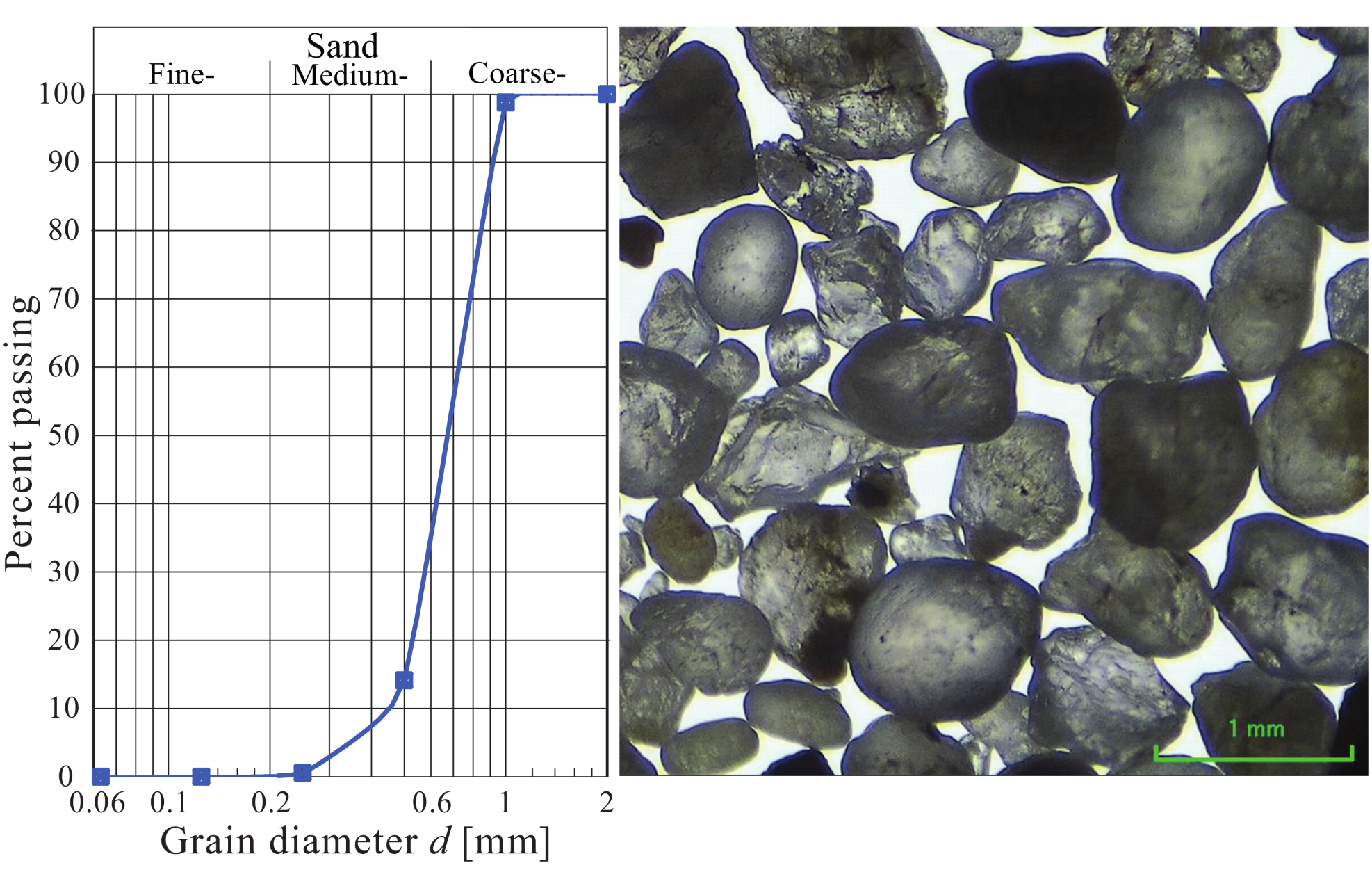}
  \end{subfigure}%

  \begin{subfigure}[b]{.9\textwidth}
    \centering
    \resizebox{0.55\linewidth}{!}{
    \begin{tabular}{c c c c c c}
        \hline
        $\rho_{s}$ & $e_{\min}$ & $e_{\max}$ & $d_{10}$ & $d_{50}$ & $d_{\max}$ \\\relax
        [g/cm$^{3}$] & [-] & [-] & [mm] & [mm] & [mm]\\
        \hline
        2.64 & 0.52 & 0.805 & 0.45 & 0.68 & 2.0\\
        \hline
    \end{tabular}
    }
  \end{subfigure}
    \caption{\textcolor{black}{Grain size distribution, a microscopic image and selected soil parameters of Hamburg sand.}}
    \label{fig:gsd_micro} 
\end{figure}

The original sand is washed over a No. 230 U.S. Standard sieve to remove the finer grain fractions with diameters close to the spatial resolution of the CT images. Furthermore, particles containing iron ore are removed with the help of a magnet, thus reducing the occurrence of metal artifacts in CT images. 
\textcolor{black}{The grain size distribution and selected soil parameters of Hamburg sand are summarized in~\cref{fig:gsd_micro}}

The cylindrical sample of 12 mm height and 12 mm diameter is placed in a flow cell. Starting from an initially saturated state, drainage and imbibition are carried out by forced outflow or inflow through a filter membrane and porous plate at the bottom of the sample. The flow is applied with a miniaturized 3D-printed syringe pump based on~\citet{wijnen2014open}. 

\textcolor{black}{During specimen preparation, dry sand is carefully pluviated into the partially water-filled flow cell and compacted in layers. To prevent segregation, each layer is roughened with the tip of a screw driver. After reaching the intended specimen height corresponding to a selected initial void ratio, a topcap with a central bore hole to connect the specimen to ambient air is placed on top of the specimen.}

\textcolor{black}{Macroscopically, this procedure leads to an initially water-saturated
specimen. However, by inspecting CT data of the initial specimen state, on the microscopic level, many small gas bubbles were found in the pore space, the smallest of
which must likely be attributed to imaging noise due to their size of single voxels or small voxel groups. The insights from CT show that from an experimental point of view, the sand pluviation technique is not
perfect if an ideally fully water-saturated specimen is to be created. Based on CT data, the initial macroscopic degree of saturation was found to be $S_r = 0.988$ \citep{Milatz_et_al2022}.}

The change in capillary pressure or matric suction is measured with pore liquid pressure sensors in the drainage system below the samples. The whole experimental setup is controlled by the Raspberry Pi 3B+ single-board computer, which controls the hydraulic paths. Python scripts are used to control the stepper motor of the syringe pump and simultaneously log matric suction as measured by the pore liquid pressure sensor. For further technical details please refer to~\citet{Milatz2020} and~\citet{Milatz_et_al2022}.

The in situ experiment consists of 20 CT scans and 19 hydraulic steps. A tomography image is obtained following each hydraulic step that is stopped at similar degrees of saturation on various hydraulic paths. The temporal change in the prescribed macroscopic degree of saturation and measured matric suction is in accordance with~\citet{Milatz_et_al2022}. 
The degree of saturation linearly changes with the flow rate, defined as the volume of liquid pumped in or out of the sample ($V_{\omega}$) per unit time ($t$), of $\frac{\partial V_{\omega }}{\partial t}=$~\SI{0.0597}{\milli\meter\cubed\per\second} corresponding to $\frac{\partial S_{r}}{\partial t} \approx $~\SI{0.000112}{\per\second}. 

At every CT scan step, the whole sample volume is captured and segmented into the corresponding phases using a multiphase image analysis tool to investigate the capillary processes, as described in~\citet{Milatz_et_al2022}.

\subsection{Multiphase LBM-simulation}
\label{subsec:multiphase_LBM_approach}

We develop a GPU-capable 3D multiphase lattice Boltzmann method (LBM) code to model the liquid-gas phase-transition in unsaturated soils. In this study, we use the D3Q19 scheme for velocity discretization~\citep{d1986lattice}. Within this scheme, the fluid domain is divided into an equally-spaced grid in every direction. This grid is called the lattice, and grid intersections are referred to as lattice nodes. D and Q denote the lattice dimension and the number of discrete velocities, respectively. 
Each lattice node at each discrete direction, $i$, has corresponding particle distribution function, $f_{i}$ , which represents the density of the particles with velocity $\boldsymbol{c}_{i}$ at the given node. Macroscopic quantities such as mass density, $\rho$, or fluid velocity, $\boldsymbol{u}$, which are usually the parameters of interest in fluid dynamics, can be calculated from the distribution functions using
\newcommand{\dd}[1]{\mathrm{d}#1}  
\begin{equation}
\rho=\sum_{i=0}^{19}f_{i},\,
\end{equation}
and
\begin{equation}
\rho\boldsymbol{u}=\sum_{i=0}^{19}f_{i}\boldsymbol{c}_{i},\,
\end{equation}
Following the definition of the velocities, an evolution rule is applied to solve the Boltzmann equation at every time increment:
\begin{equation}
f_{i}(\boldsymbol{x}+\boldsymbol{c}_{i}\Delta t,t+\Delta t)=f_{i}(\boldsymbol{x},t)+\mathrm{\Omega}_{col},\,
\end{equation}
where $\boldsymbol{x}$ is the position of the lattice node, $t$ is the current time and $\mathrm{\Omega}_{col}$ is the Bhatnagar–Gross–Krook (BGK) collision operator~\citep{d1986lattice}:
\begin{equation}
\mathrm{\Omega}_{col}=\frac{f_{i}-f_{i}^{eq}}{\tau }\Delta t,\,
\label{eq:bgk_operator}
\end{equation}
During collision, the collision operator depends on the relaxation time $\tau$ which we use ${\tau = \Delta t}$ in our simulations. The equilibrium distribution function, $f_{i}^{eq}$ , in~\cref{eq:bgk_operator} is given by,
\begin{equation}
f_{i}^{eq}=\omega _{i}\rho (1+\frac{\boldsymbol{u}\cdot \boldsymbol{c}_{i}}{\boldsymbol{c}_{s}^{2}}+\frac{(\boldsymbol{u}\cdot \boldsymbol{c}_{i})^2}{2\boldsymbol{c}_{s}^{4}}-\frac{\boldsymbol{u}\cdot \boldsymbol{u}}{2\boldsymbol{c}_{s}^{2}}),\,
\end{equation}
where $\omega _i$ as the corresponding weights. After a collision, the updated distribution functions are streamed according on their velocities: each $f_i$ is pushed one lattice node in the $i$ direction.

We introduce the Shan-Chen (SC) body forces to achieve the phase transition between liquid and gas nodes and to model adhesion. The adhesion force between the liquid/gas phase and solid walls is calculated as
\begin{equation}
F_{SC}(\boldsymbol{x})=-\psi (\boldsymbol{x})G\sum_{i}\omega _{i}\psi (\rho)c_{i}\Delta t,\,
\end{equation}  
where $\psi$ is an effective density, $G$ is a parameter that controls the repulsion intensity (positive for repulsion and negative for attraction. 
The interaction between the solid and fluid nodes is modeled using a bounce-back boundary condition \citep{Huang2015,kruger2017lattice}. At bounce-back, the fluid is repelled by the solid upon collision and propagates the particle distribution to the neighboring fluid domain. The SC body forces are calculated based on the post-collision particle distribution and used to update the equilibrium velocity before streaming. The SC body forces are given as
\begin{equation}
F_{SC}(\boldsymbol{x})=-\psi (\boldsymbol{x})G\sum_{i}\omega _{i}\psi (\boldsymbol{x}+c_{i}\Delta t)c_{i}\Delta t,
\end{equation} 
Since $\psi$ is a function of density and pressure, an Equation of State (EOS) is required to define the dependence of pressure on density. We employ the Carnahan-Starling (C-S) EOS similar to~\citet{hosseini2022investigating} and obtain the same coexistence curves at different temperatures $T$. The threshold densities defines the boundary that separates liquid ($l$) densities above and gas ($g$) densities below. We use a ${\rho}_{l} = 0.2725~\mathrm{(mu)\cdot {(lu)}^{-3}}$ and ${\rho}_{g} = 0.0484~\mathrm{(mu)\cdot {(lu)}^{-3}}$ in the multiphase simulation to identify liquid nodes and gas nodes, where mu and lu are the mass and length units in lattice Boltzmann methods. The values of ${\rho}_{l}$ and ${\rho}_{g}$ are selected at which first derivatives of the EOS are zero. Fluid densities between ${\rho}_{l}$ and ${\rho}_{g}$ define a transition zone. The solid ($s$) density $\rho_{s}$, also known as wall density, is set to ${\rho}_{s} = 0.35~\mathrm{(mu)\cdot (lu)^{-3}}$, corresponding to a constant contact angle of $5^{\circ}$, which means nearly perfect wetting.

\begin{figure}[htbp]
\begin{center}
\includegraphics[width=0.6\columnwidth]{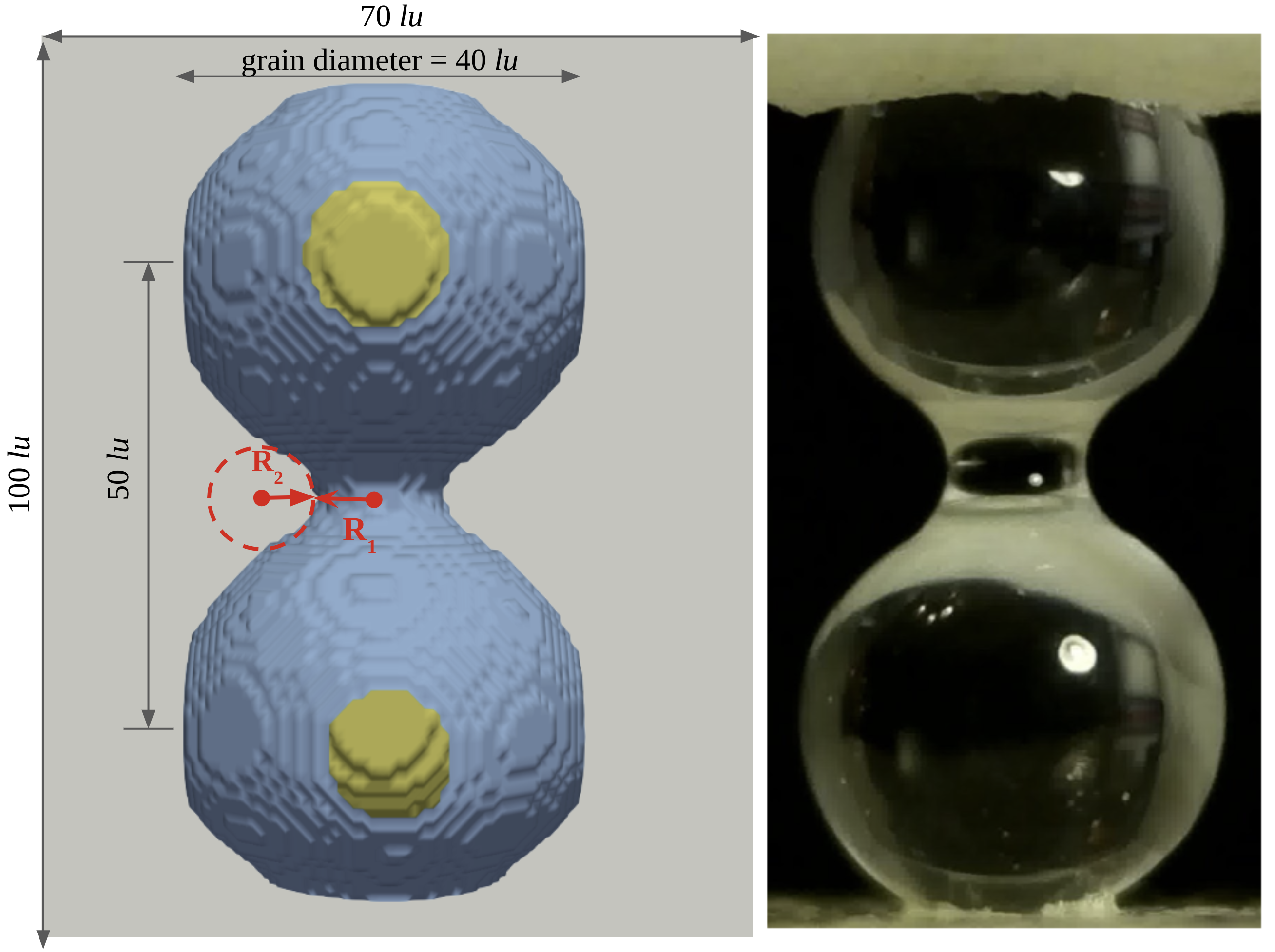}
\end{center}
\caption{Comparison of numerically simulated (left) and experimentally (right) observed capillary bridges The left figure shows the liquid (blue) distribution after 16500 time steps of phase separation. $R_{1}$ and $R_{2}$ show the principal radii of curvature of the liquid-gas interface.}
\label{Fig:observed_capillary_bridges}
\end{figure}

In LBM, the spatial and time domains are scaled using lattice units. To accurately capture the real physics, we convert quantities between lattice units and physical units. Consequently, we employ the Young-Laplace-equation, which describes the relationship between capillary pressure and surface tension for a capillary meniscus as
\begin{equation}
p_{c}=\gamma (\frac{1}{R_{1}}+\frac{1}{R_{2}}),
\label{eq:Young-Laplace}
\end{equation} 
where $p_{c}$ is the capillary pressure, $\gamma$ is the surface tension of liquid at $20 ^{\circ}\mathrm{C}$, and $R_{1}$ and $R_{2}$ are the principal radii of curvature of the liquid-gas interface. To derive $\gamma$ in lattice units, we set up a simple configuration of a two-grain model to measure the geometry of the capillary bridge as shown in~\cref{Fig:observed_capillary_bridges}. We set up two immobile grains vertically with centers 50 lu apart and radii of 20 lu inside a periodic fluid domain with dimensions of 100 lu $\times 70$ lu $\times 70$ lu. We initialize the pore space with a small amount of liquid ($S_{r}=0.02$) to form a capillary bridge at equilibrium. We then slowly drain the liquid at a rate of 0.01 lu every 1000 time steps, until a capillary bridge forms at a saturation of 0.004. The suction stabilizes at $p_{c}=0.0041~\mathrm{(mu)\cdot (lu)^{-1}(ts)^{-2}}$ 
((ts) is the time unit in lattice Boltzmann methods). We measure $R_{1}$ and $R_{2}$ along the saddle surface at the neck of the capillary bridge, shown as the red circle in~\cref{Fig:observed_capillary_bridges}. $R_{1}$, $R_{2}$, and the contact angle are measured manually from the 2D projection of the capillary bridge. Conventionally, $R_{1}$ is measured orthogonal to the meridional profile of the capillary bridge so that its center of curvature is along the bridge's axis of symmetry. So, the center of curvature of $R_{1}$ is always inside the liquid bridge, and $R_{1}$ itself positive and calculated at the neck. $R_{2}$ is measured along a meridional bridge profile and can be positive or negative depending on whether its center of curvature is inside or outside the liquid bridge \citep{schubert1982kapillaritat,willett2000capillary}. We obtain $R_{1} = 5.83$ lu and $R_{2} = -6.56$ lu given the saddle-shaped liquid-gas interface.

The surface tension is then derived using~\cref{eq:Young-Laplace} as $2.148\times10^{-2}~\mathrm{(mu)\cdot (ts)^{-2}}$. We normalize the suction from both CT and LBM using ${\gamma}/{d_{50}}$, where $d_{50}$ is a is the representative mean grain diameter of Hamburg sand. For CT experiment, we measure a surface tension of $\gamma=$~\SI{0.07275}{\newton\per\meter} and $d_{50} =$~\SI{0.68}{\milli\meter} as determined from the grain size distribution curve. In LBM, $d_{50} =\frac{ \SI{0.68}{\milli\meter}}{\Delta x}$ where $\Delta x$ is the physical length per lattice unit. In this study, the physical domain is modeled using 400 LB nodes in each dimension, so $\Delta x = \frac{\SI{8}{\milli\meter}}{400} = 2\times10^{-5}{\si{\meter}}\cdot\mathrm{(lu)^{-1}}$ and $d_{50} =34~\mathrm{(lu)}$. Thus, the normalized surface tension in the CT experiment is ${\gamma}/{d_{50}}=$~\SI{0.107}{\kilo\pascal} and in LBM simulation is ${\gamma}/{d_{50}}= 6.318\times10^{-4}~\mathrm{(mu)\cdot (lu)^{-1}(ts)^{-2}}$. 

\begin{figure}[htbp]
\begin{center}
\includegraphics[width=\columnwidth]{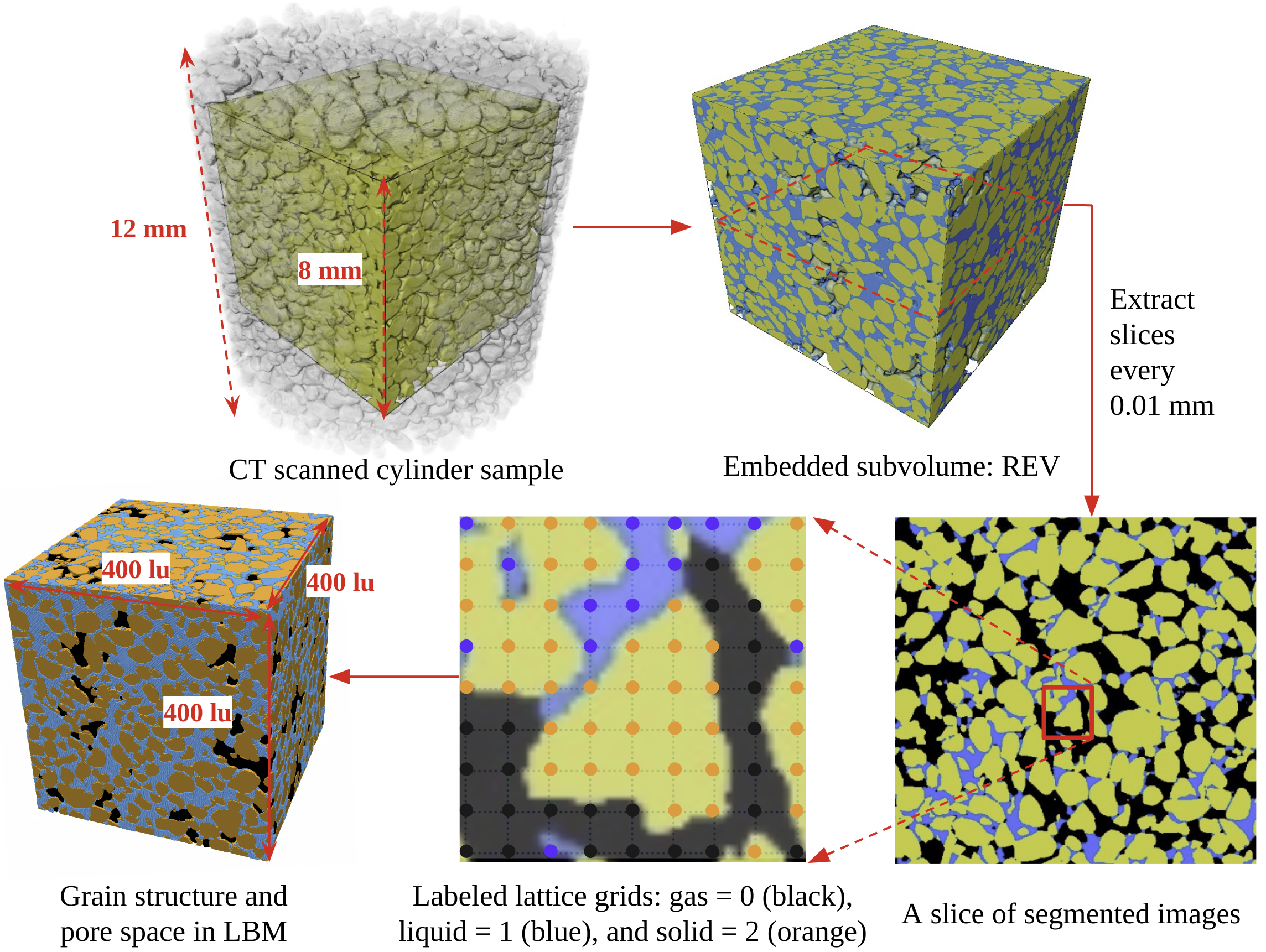}
\end{center}
\caption{\textcolor{black}{Schematic diagram of constructing grain structure and pore space in LBM based on CT-scan images.}}
\label{Fig:Conversion_CT_LBM}
\end{figure}

To construct a digital twin of the CT experiment using LBM, we perform phase segmentation on the scanned 12-mm-diameter cylinder sample from the CT experiment. The segmented data is stored as Tag Image File Format (TIFF) which consists of 1200 slices of images showing the cross section of the cylinder sample every 0.01 mm throughout its height. Every slice represents a labeled volume with all voxels identified as a separate phase and labeled with integer values (gas = 0, liquid = 1, and solid = 2). We then select a sub-volume of 8-mm-high central cube from the cylinder sample as the representative elementary volume (REV), which contains 2937 grains. This cube is interpreted as a 3D matrix of solid grain and pore space voxels filled with liquid and gas. We use the 3D matrix to construct the grain structure in the LBM-model. The LBM uses a step-wise approximation to construct the rough irregular solid boundaries; this introduces a small error in the model. We use a very fine resolution of 400 x 400 x 400 to minimize this discretization error.~\Cref{Fig:Conversion_CT_LBM} illustrates the construction of the LBM model from the CT experiment in detail. We use a scale factor of 2, so the grid size is $400~\mathrm{lu}\times 400~\mathrm{lu}\times400~\mathrm{lu}$. Scaling in the LBM simulation is analogous to binning in image processing. 

After initializing the grain structure in LBM, we randomly initialize the fluid densities throughout the pore space to a value in the transition zone, with a small perturbation introduced to allow the phase separation. The LBM simulation is a digital twin of the CT experiment on Hamburg sand. To eliminate boundary effects, the periodic flow boundary conditions are adopted in each direction and the grains are assumed to be immobile. Following phase separation and equilibrium, we perform uniform drainage and imbibition in the entire pore space. In the CT experiment, the injection and drainage are done through the bottom porous plate. However, we simulate a uniform injection and drainage at a constant flow rate of $7200~\mathrm{(mu)\cdot (lu)^{-3}}$ by varying the amount of density changed at each liquid node for every 5000 steps. At $S_{r} = 0.30$ during imbibition, for instance, we increase the density at each of the 7.2 million liquid nodes by $0.001~\mathrm{(mu)\cdot (lu)^{-3}}$. 

\subsection{{Micromechanical parameters for multiphase analysis
}}
\label{subsec:LBM-simulations}

In LBM, the WRC is computed by measuring the average suction ($s$) of all fluid nodes after a fixed number of time steps after varying the liquid density. The suction is calculated as the difference between the average gas pressure and the average liquid pressure across the fluid domain. Degree of saturation ($S_{r}$) is the number of liquid nodes divided by the number of fluid nodes (voids), consistent with the degree of saturation defined in soil mechanics. 

We consider various capillary state variables and pore-scale measures to statistically describe the multiphase system. The capillary state variables are microscale quantities computed from either CT images or LBM results. The capillary state variables include geometric characteristics such as liquid or gas clusters and their distribution, and interfacial areas between different phases, \ie liquid and solid, liquid and gas. 

The focus of this contribution is on the evolution of the capillary state variables in fluid clusters and interfaces along different hydraulic paths. We calculate the interfacial areas between the liquid and gas, liquid and solid, and solid and gas, as these interfaces may be used to calculate effective stress in unsaturated granular media.
\textcolor{black}{To calculate the area of the interface between different phases, we use three counters for each of the gas-liquid, liquid-solid, and solid-gas interfaces. After each injection period, we evaluate the phase of all its adjacent nodes. If an adjacent node has a different phase than the current fluid node under consideration, then we increment the counter of the corresponding interface area.}

The volume of liquid and gas clusters can be determined by counting the number of liquid and gas nodes within the computational domain or voxels within the REV of the CT data set. We identify the number of liquid and gas clusters, $N_{lc}$ and $N_{gc}$, at each injection step. 

Apart from fluid cluster statistics, we also calculate
the interfacial area between solid-liquid ($A_{sl}$), which reflect the wettability of grains, and gas-liquid ($A_{gl}$) that represents surface of liquid menisci. We normalize both the interface areas by the solid surface area ($A_{solid}=9,714,944$ nodes). We compute the capillary state variables using the same procedures to compare numerical and experimental results directly. For the CT experiment, we read 20 trinarized tomography images to extract the 3D multiphase distribution at different constant degrees of saturation. For LBM simulation, we compute the 3D multiphase distribution every 5000 steps.

\textcolor{black}{To reproduce the fluid statistics in the CT experiments, we read the phase-segmented CT images to capture the spatial distribution of liquid and gas nodes within the REV. Based on this distribution, we compute the saturation within the REV. Following that, we compare the fluid cluster statistics and interfacial areas at the same saturation using the Depth-First Search (DFS) algorithm as in the LBM simulation~\citep{tarjan1972depth}.}

Finally, we consider how the pore size affects the pore emptying and pore filling processes using a sphere-placement (chamber size) method to define the pore size \citep{sweeney2003pore}. At each fluid node, we place a sphere with a radius equal to the distance to the nearest solid node and store its value. As we iterate through all the fluid nodes, placing a sphere at each node, we update the radius value of all other fluid nodes, which are within the sphere, if their radii are smaller than the radius of the current sphere. This approach updates all the fluid nodes in a pore, by the radius of the largest sphere. We finally segment the pores as groups of fluid nodes with the same radius. Here, the sphere radius is analogous to the individual pore size.

\begin{figure}[t!]
  \centering
  \begin{subfigure}[t]{0.48\columnwidth}
    \centering
    \adjincludegraphics[width=\columnwidth]{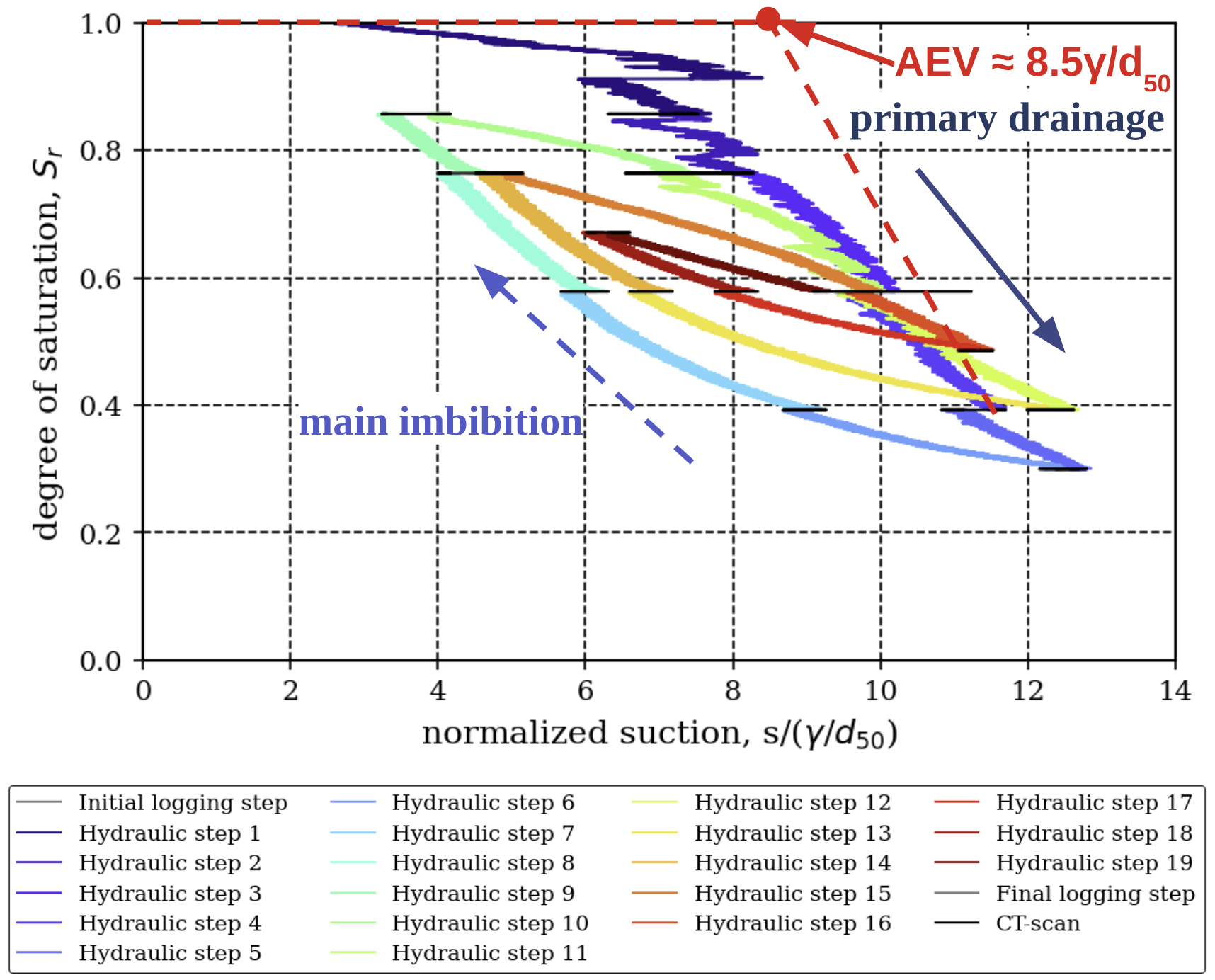}
    \caption{CT experiment}
  \end{subfigure}%
  ~ 
  \begin{subfigure}[t]{0.45\columnwidth}
    \centering
    \adjincludegraphics[width=\columnwidth]{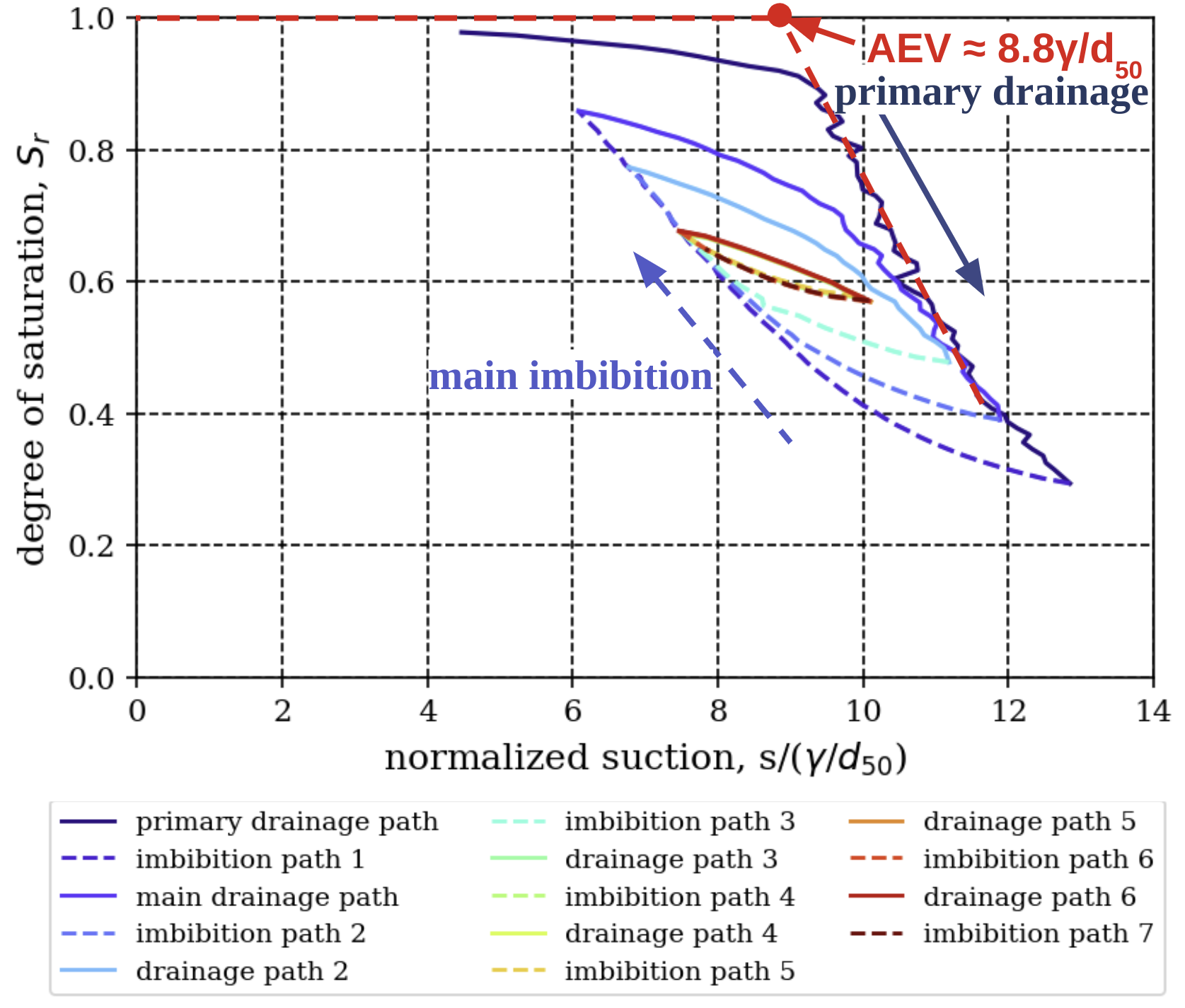}
    \caption{LBM simulation}
  \end{subfigure}
 
\caption{Cyclic WRC measured in the water retention test on Hamburg sand with CT imaging (a) and from LBM simulation (b).}
\label{Fig:comparison_of_wrcs}
\end{figure}

\section{Results}
\label{sec:results}

\subsection{Water retention curves from CT and LBM}

\label{subsec:macroscopic_liquid_retention curves}

\Cref{Fig:comparison_of_wrcs} shows the water retention curve for consecutive hydraulic steps in CT imaging and LBM simulation. Both the LBM and CT show similar primary drainage paths and Air Entry Value (AEV) of $8\sim9\,\gamma/d_{50}$. The LBM simulation shows a fairly smooth change at the AEV, while CT experiment show a sudden drop in suction value, probably due to the formation of localized gas clusters in the vicinity of the filter plate breaking the hydraulic connection to the pressure sensor. The suction drop makes it harder to get an accurate estimate of suction the experiment. At $S_{r}=0.3$, we observe a maximum suction in the LBM simulation of 12.7 $\gamma/d_{50}$, which matches the measured suction of 12.9 $\gamma/d_{50}$ in the CT experiment.

There are also noteworthy differences between CT and LBM results. Along the entire main imbibition path, the LBM predicts higher suction than the CT experiment for the same degree of saturation. At the end of main imbibition $S_{r}=0.83$, the LBM yields a higher suction $s=6\gamma/d_{50}$ in comparison to $s=4\gamma/d_{50}$ for the CT experiment.

The CT experiment captures the transient response while LBM allows sufficient time between saturation change to equilibrate the pore pressures. The transient WRC response results in a relaxation of suction during scans, yielding a lower suction along imbibition paths~\citep{milatz2018theoretical,galindo2013lattice}. 

In LBM simulations, the primary drainage and main imbibition paths enclose the subsequent imbibition and drainage scanning paths. Although the WRC from the CT experiment show similar trends, the maximum suction in subsequent drainage paths overshoots the primary drainage path. The LBM simulations capture the equilibrium response of WRC using a rest interval of 5000 steps after density changes. Whereas the CT experiment measures the transient response without allowing the system to equilibrate, resulting in an overshooting of suction at the end of scanning drainage paths. Differences in suction measurement and sample preparation between LBM and CT could also affect the overshooting of suction in scanning drainage paths.

In CT experiment, suction is measured at the bottom of the sample, where the local distribution of gas clusters may result in localized suction estimates, which maybe different from global suction values. In LBM simulation, suction is continuously computed as the difference between the average liquid and gas pressures in the entire sample. The CT sample is prepared by dry sand pluviation, the liquid level within the sample holder is kept above the top of the sample without applying back pressure to exclude the gas in the sample. The CT data, therefore, shows thousands of initial air entrapment in the pore space, a portion of which, especially the small clusters, might also be related to image noise. The entrapped gas bubbles (see solid-triangle zones in~\cref{Fig:gas_cluster_distribution_2Dslices_drainage}a) could result in an inaccurate suction measurement as compared to the LBM simulation, which starts from a state without entrapped gas.

\subsubsection{Spatial distribution of liquid and gas phases during drainage and imbibition}

\begin{figure*}[htbp]
\begin{center}
\begin{subfigure}{\columnwidth}
  \includegraphics[width=\textwidth]{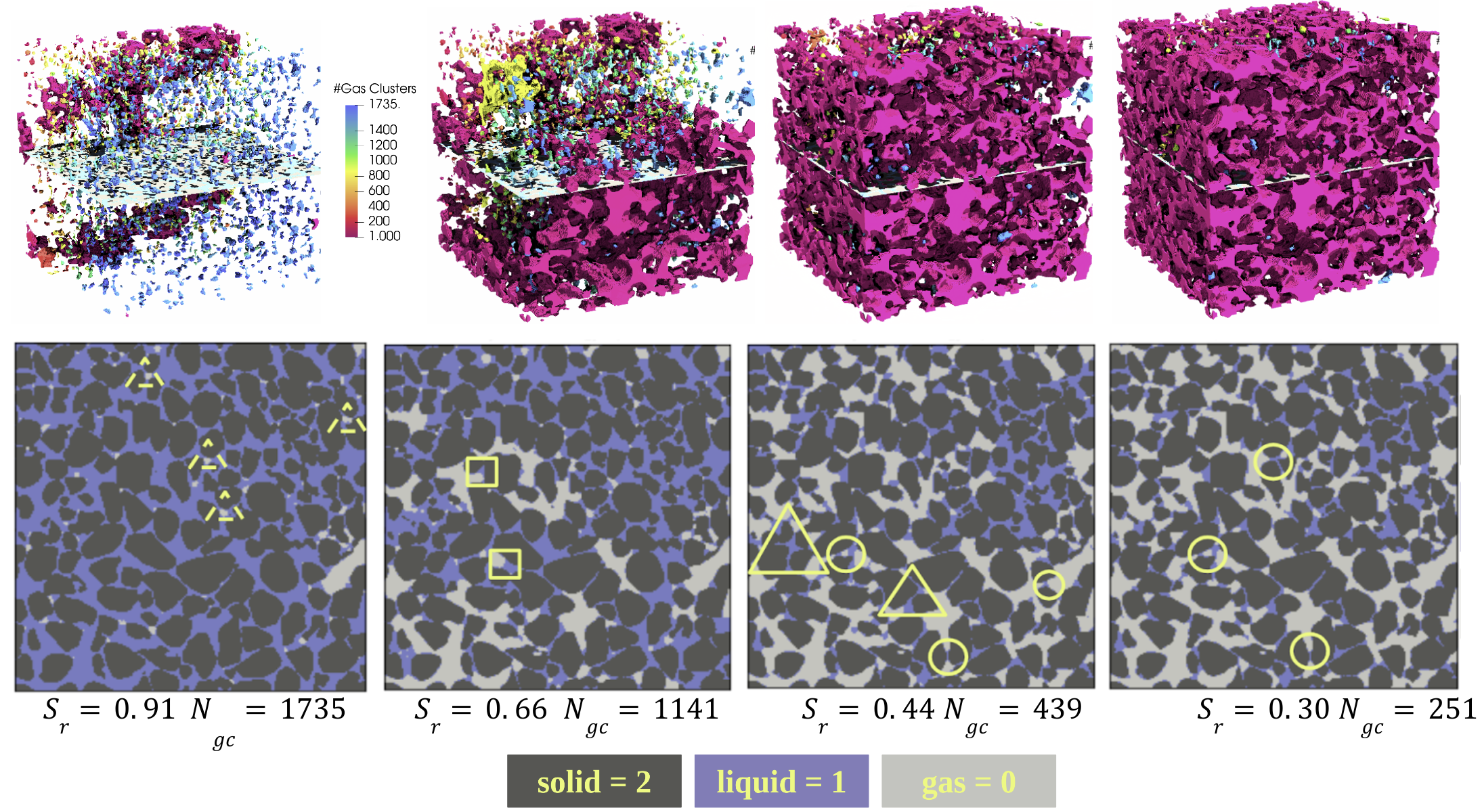}
\caption{Gas cluster distribution with 2D slices of the CT experiment}
\end{subfigure}
\begin{subfigure}{\columnwidth}
  \includegraphics[width=\textwidth]{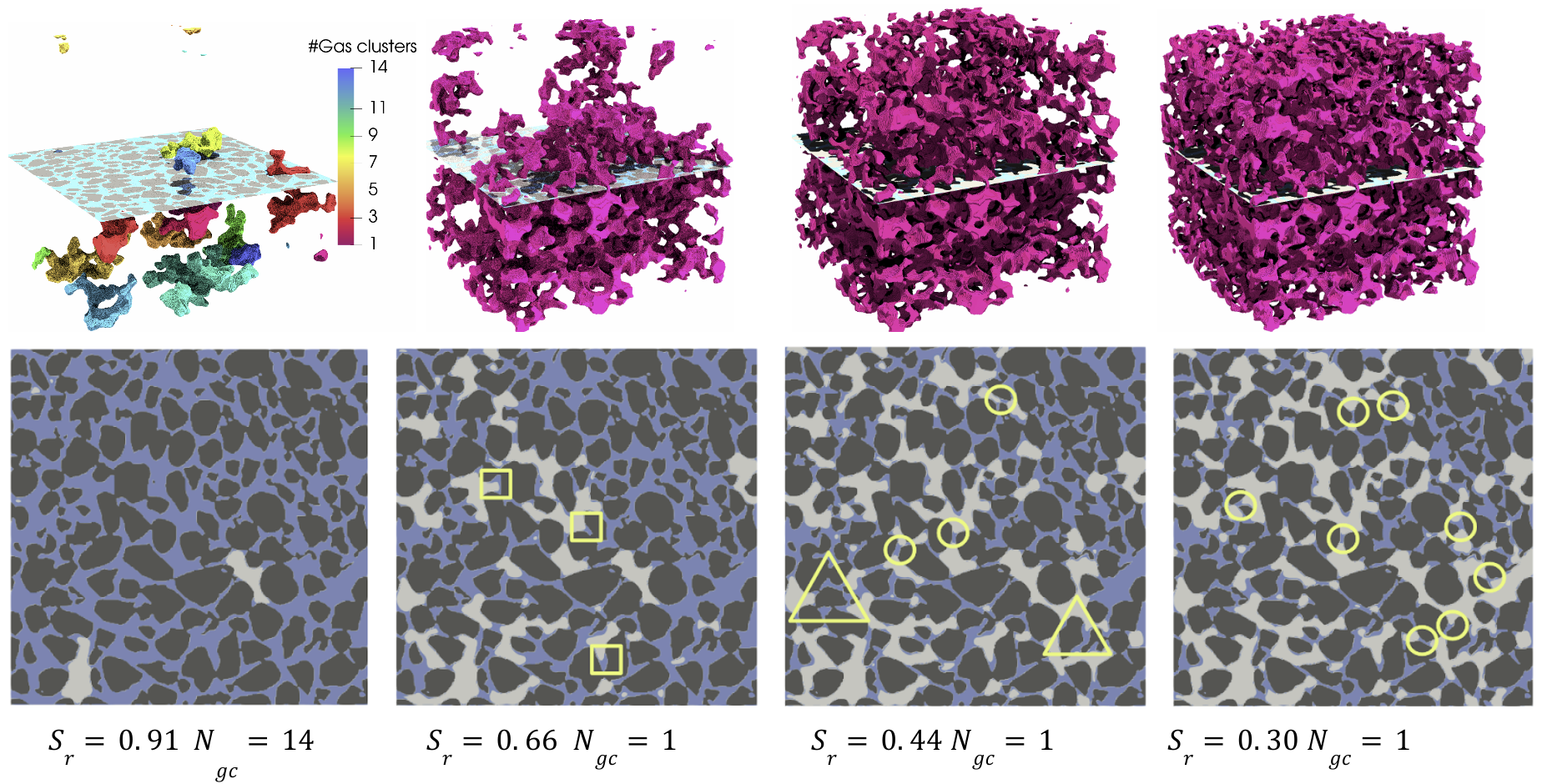}
\caption{Gas cluster distribution with 2D slices of the LBM simulation}
\end{subfigure}
\caption{3D distribution of gas clusters as well as snapshots of 2D slices of liquid (blue) and gas (white) phases at mid-height of the sample at different $S_{r}$ along the primary drainage path. The dashed-triangle zones show examples of the initial air entrapment in CT experiment in the funicular regime; the solid-triangle zones, solid-square zones, and solid-circle zones show examples of a liquid cluster connecting multiple grains, examples of concave menisci, and examples of liquid bridges in the capillary regime.}
\label{Fig:gas_cluster_distribution_2Dslices_drainage}
\end{center}
\end{figure*}

\begin{figure*}[htbp]
\begin{center}
\begin{subfigure}{\columnwidth}
  \includegraphics[width=\textwidth]{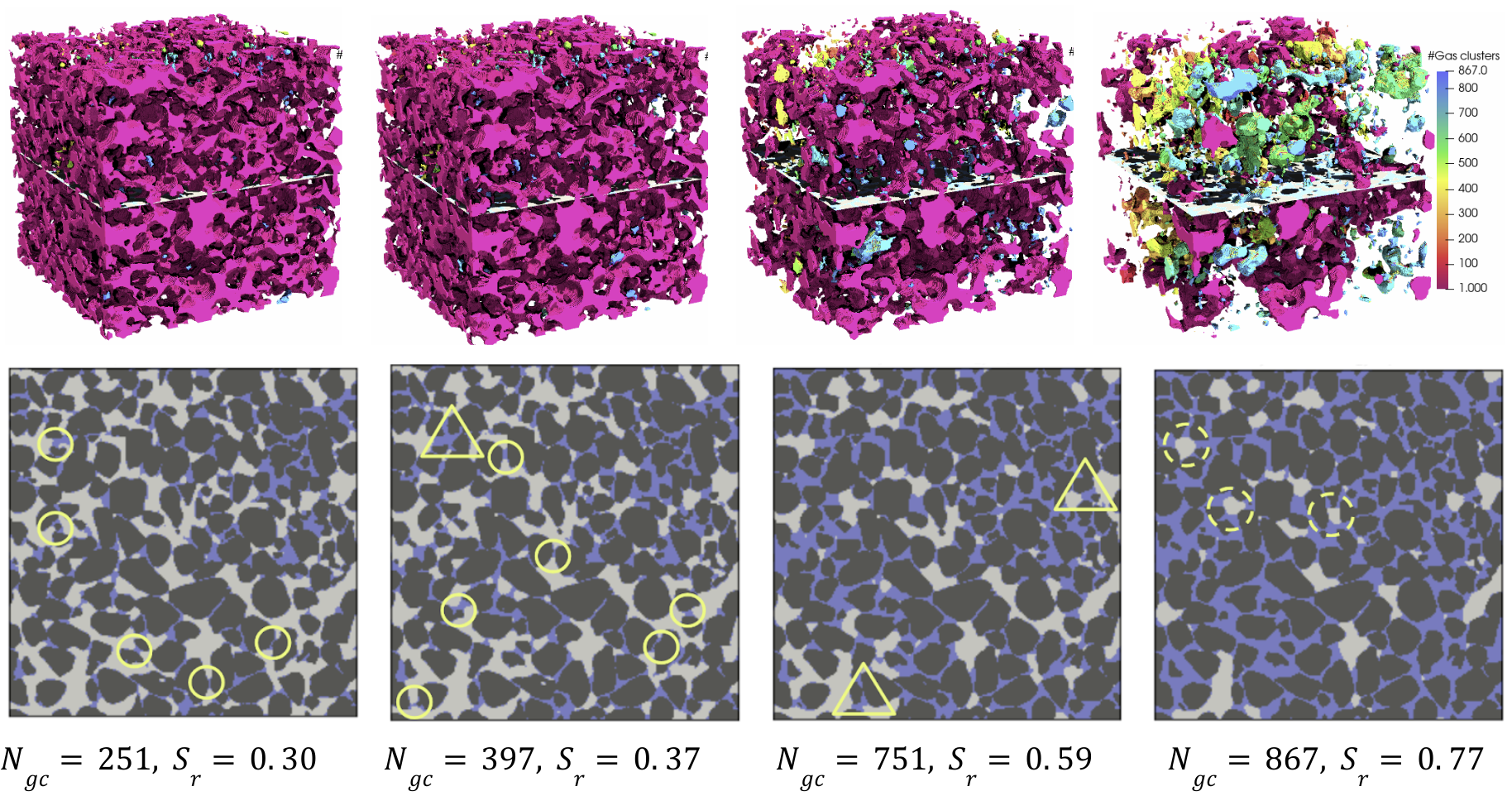}
\caption{Gas cluster distribution with 2D slices of the CT experiment}
\end{subfigure}
\begin{subfigure}{\columnwidth}
  \includegraphics[width=\textwidth]{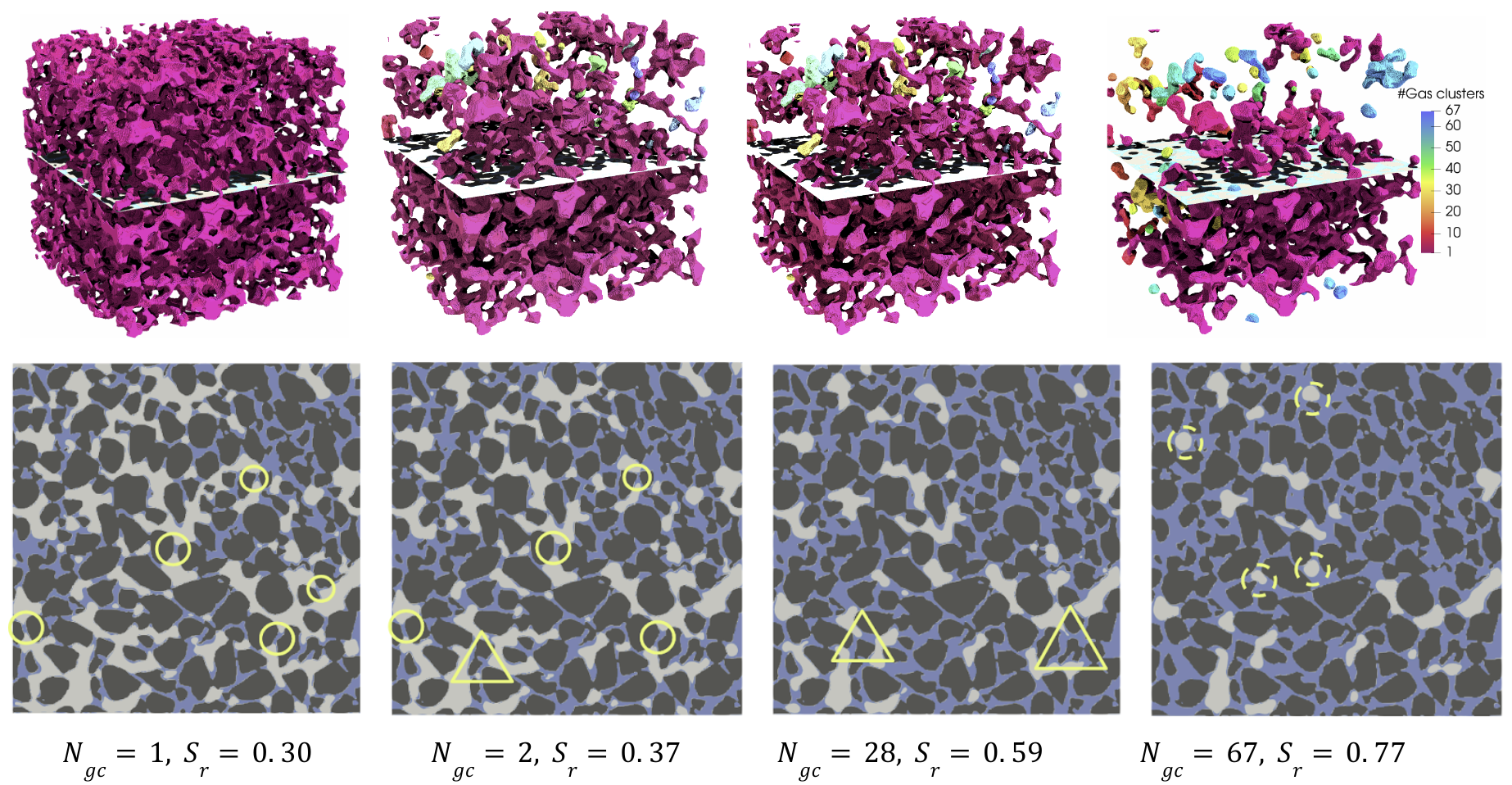}
\caption{Gas cluster distribution with 2D slices of the LBM simulation}
\end{subfigure}
\caption{3D distribution of separate gas clusters as well as snapshots of 2D slices of liquid (blue) and gas (white) phases at mid-height of the sample at different $S_{r}$ along the main imbibition path. The solid-circle zones show examples of binary liquid bridges, the solid-triangle zone shows examples of liquid clusters connecting multiple grains in the capillary state, and the dashed-circle zones show examples of individual gas bubbles in the funicular state.}
\label{Fig:gas_cluster_distribution_2Dslices_imbibition}
\end{center}
\end{figure*}
\Cref{Fig:gas_cluster_distribution_2Dslices_drainage} shows the distribution of gas clusters in the CT experiment and LBM simulation along the primary drainage. At the start of primary drainage, for the same saturation ($S_{r}=0.91$), the CT experiment shows a large number of gas clusters ($N_{gc}=1735$), while the LBM simulation only shows 14 gas clusters. This discrepancy in the number of gas clusters may be attributed to the initial air entrapment during sample preparation (see dashed-triangle zones) and imaging noise. 
In addition, the gas clusters in CT and LBM originate at different locations. In the LBM, gas clusters form in the lower half of the sample, whereas in CT, a large gas channel branches vertically and other gas bubbles appear throughout the pore space. In both LBM and CT, a single large liquid cluster occupies 90\% of the total pore space. 

As saturation decreases from 0.91 to 0.66, the liquid transits from the capillary regime where a few gas bubbles are entrapped, to the funicular regime where the gas clusters connect through pores forming its concave menisci connecting solid grains (see solid-square zones in~\cref{Fig:gas_cluster_distribution_2Dslices_drainage}). The number of gas clusters decreases with decreasing saturation as multiple gas clusters converge to form a single continuous gas cluster. At $S_r=0.3$, the majority of liquid phase reaches the pendular regime with an abundance of liquid bridges (see solid-circles zones in~\cref{Fig:gas_cluster_distribution_2Dslices_drainage}).

\cref{Fig:gas_cluster_distribution_2Dslices_imbibition} shows the distribution of gas clusters in CT and LBM along the main imbibition. We start the main imbibition by increasing the liquid density throughout the sample from $S_{r} = 0.3$. When saturation increases, existing liquid bridges expands (shown as solid circles in\cref{Fig:gas_cluster_distribution_2Dslices_drainage}) separating the continuous gas phase into multiple gas clusters. At $S_{r} = 0.59$, the majority of liquid bridges transforms into liquid clusters connecting multiple grains (shown as solid-triangle zones in~\cref{Fig:gas_cluster_distribution_2Dslices_imbibition}). As the saturation reaches $S_{r} = 0.77$, we observe the transition from a funicular to a capillary state, in which the continuous liquid phase entraps individual gas bubbles (shown as dashed-circle zones).


\begin{figure*}[htbp]
  \centering
  \begin{subfigure}[t]{0.48\columnwidth}
    \centering
    \includegraphics[width=\columnwidth]{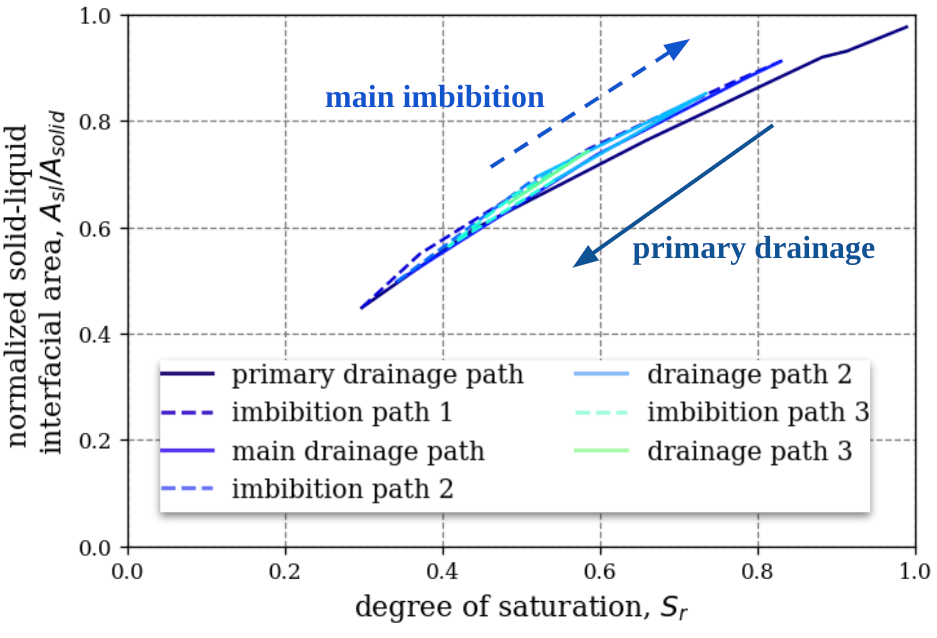}
    \caption{CT experiment}
  \end{subfigure}%
  ~ 
  \begin{subfigure}[t]{0.48\columnwidth}
    \centering
    \adjincludegraphics[width=\columnwidth]{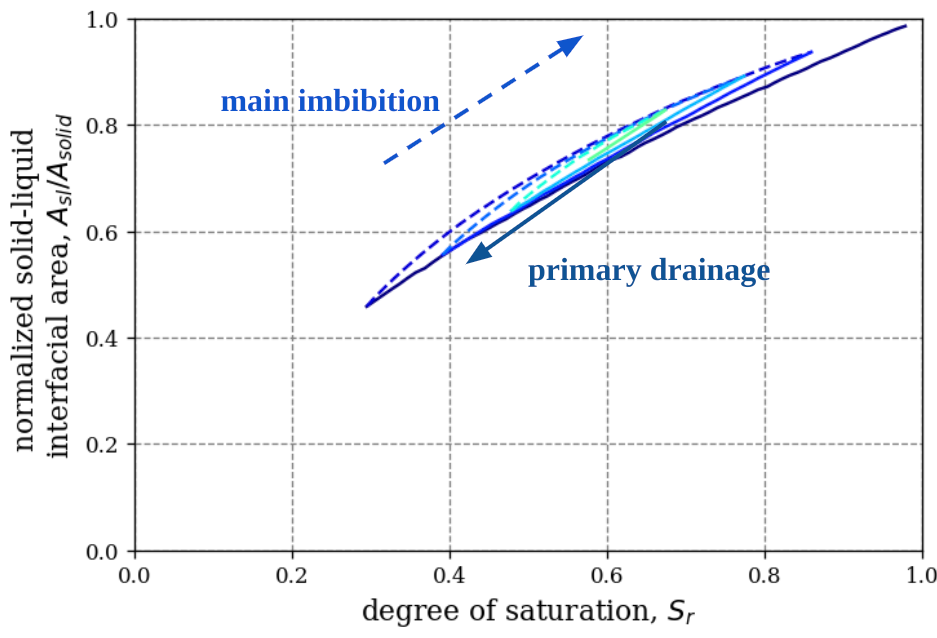}
    \caption{LBM simulation}
  \end{subfigure}%
  
  \begin{subfigure}[t]{0.48\columnwidth}
    \centering
    \includegraphics[width=\columnwidth]{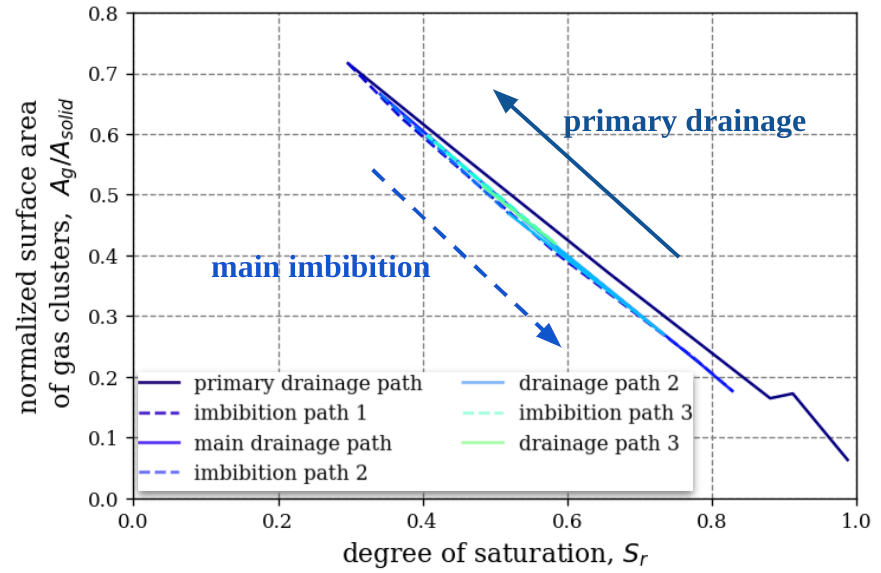}
    \caption{CT experiment}
  \end{subfigure}%
  ~ 
  \begin{subfigure}[t]{0.48\columnwidth}
    \centering
    \adjincludegraphics[width=\columnwidth]{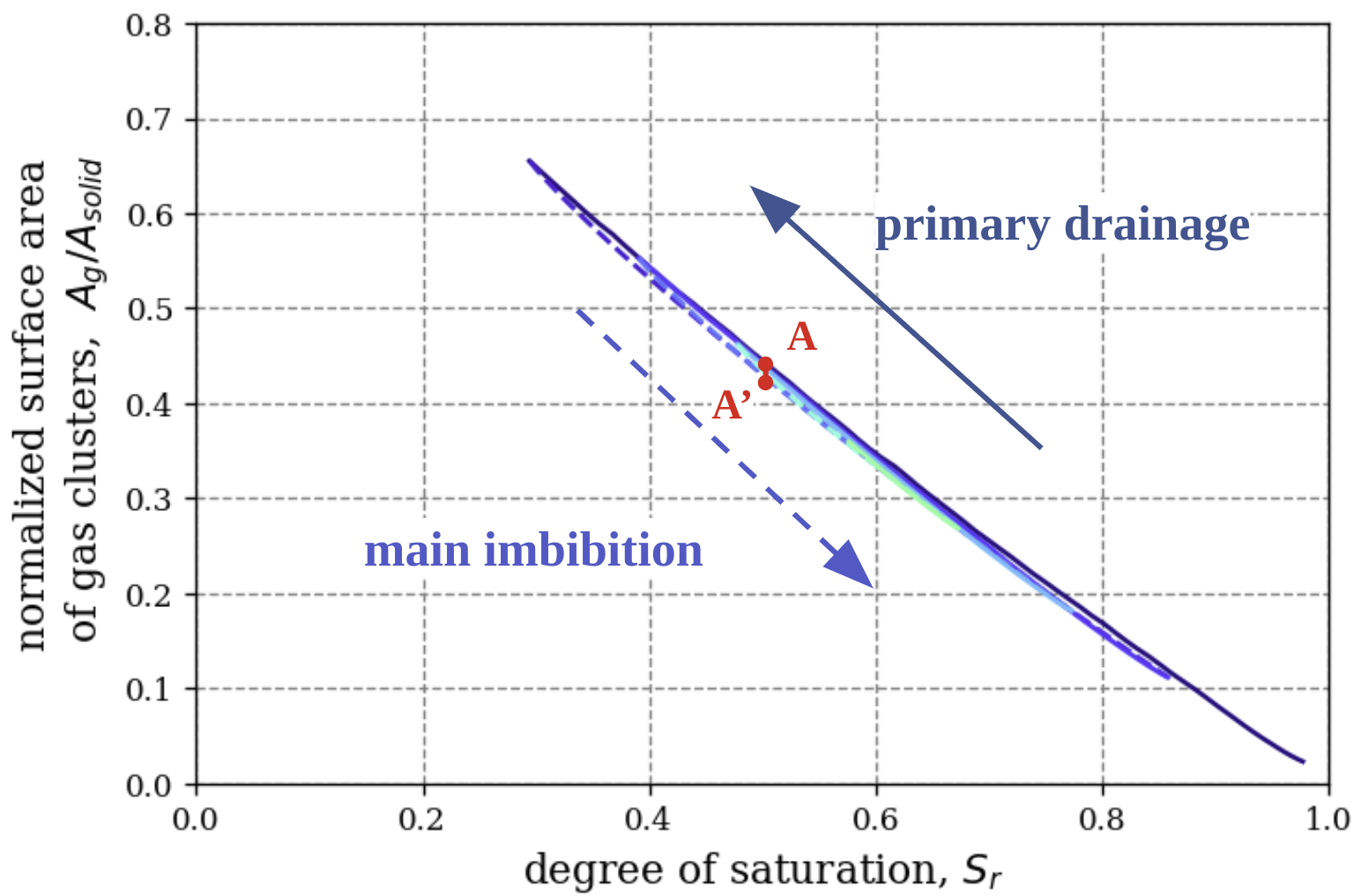}
    \caption{LBM simulation}
  \end{subfigure}
  
    \begin{subfigure}[t]{0.48\columnwidth}
    \centering
    \includegraphics[width=\columnwidth]{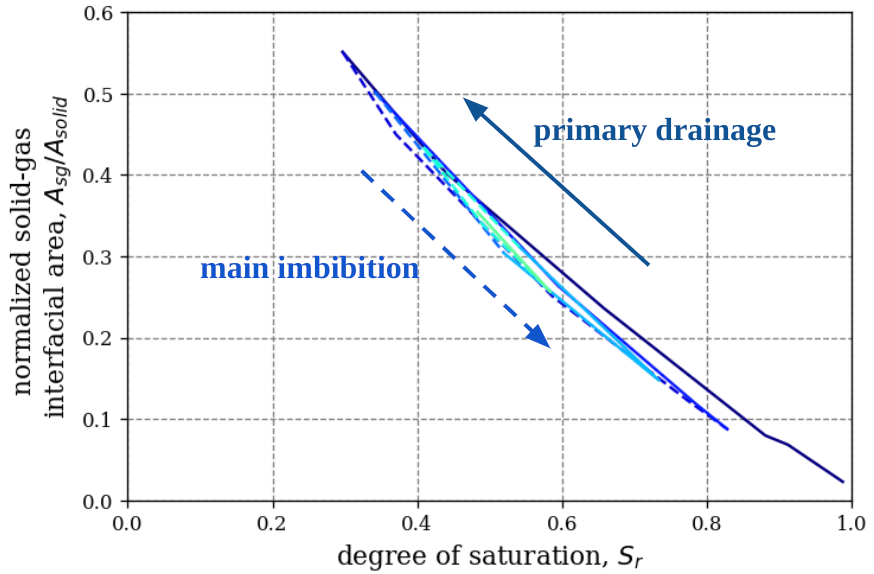}
    \caption{CT experiment}
  \end{subfigure}%
  ~ 
  \begin{subfigure}[t]{0.48\columnwidth}
    \centering
    \adjincludegraphics[width=\columnwidth]{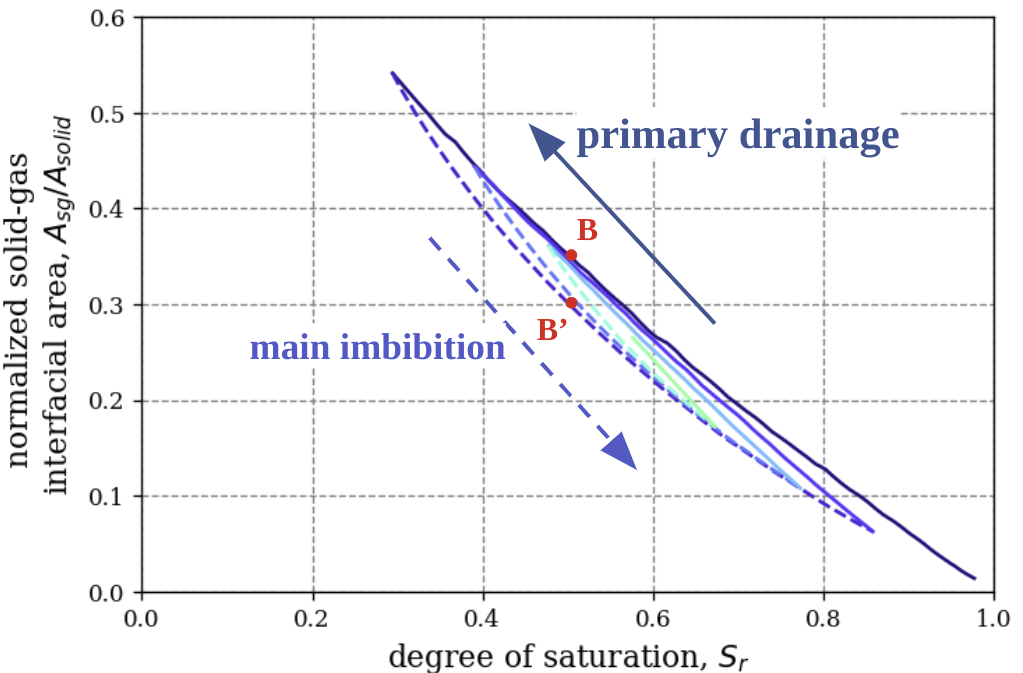}
    \caption{LBM simulation}
  \end{subfigure}
  
    \begin{subfigure}[t]{0.48\columnwidth}
    \centering
    \includegraphics[width=\columnwidth]{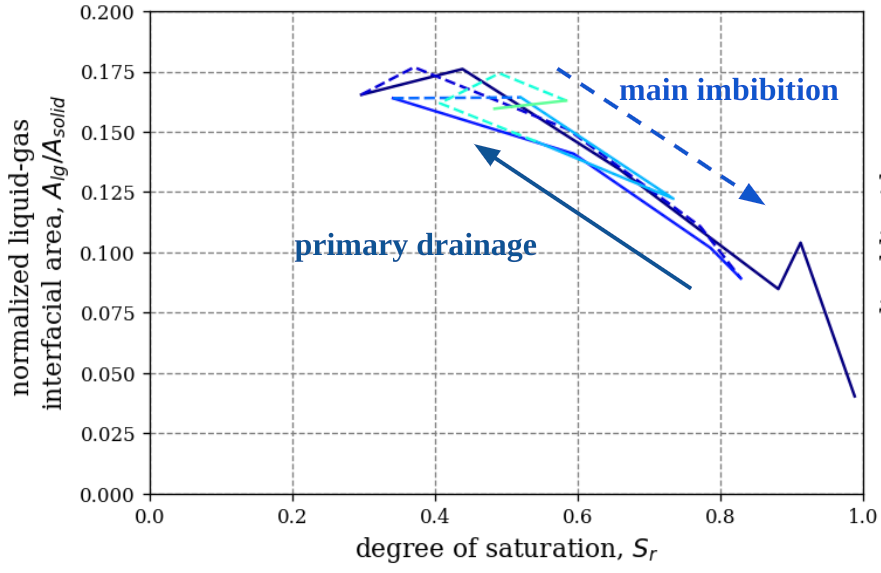}
    \caption{CT experiment}
  \end{subfigure}%
  ~ 
  \begin{subfigure}[t]{0.48\columnwidth}
    \centering
    \adjincludegraphics[width=\columnwidth]{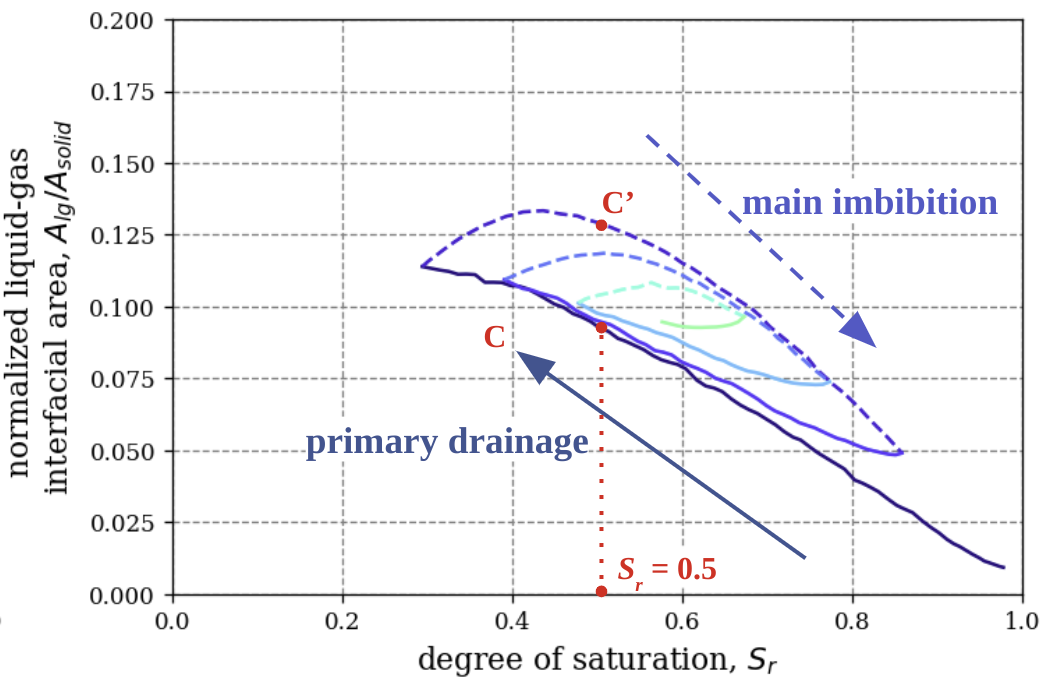}
    \caption{LBM simulation}
  \end{subfigure}
  
  \caption{Evolution of solid-liquid interfacial area ($A_{sl}$), gas surface area ($A_g$), solid-gas interfacial area ($A_{sl}$), and liquid-gas interfacial area ($A_{lg}$), normalized by the surface area of solid ($A_{solid}$) versus degree of saturation during drainage and imbibition paths in CT experiment and LBM simulation.}
\label{Fig:interfaces}
\end{figure*}

\subsubsection{Interfacial contact between liquid, gas, and solid phases}
\label{subsubsec:Interfacial contact between liquid, gas, and solid phases}

We investigate the hysteresis by examining the evolution of capillary state variables that describe interfacial contact between solid, liquid, and gas phases along drainage and imbibition paths (see definition in~\cref{subsec:LBM-simulations}).

\cref{Fig:interfaces} shows the evolution of the solid-liquid interfacial area ($A_{sl}$), surface area of gas clusters ($A_{g}$), liquid-gas interfacial area ($A_{lg}$) and solid-gas interfacial area ($A_{sg}$) all normalized by the surface area of grains ($A_{solid}$) with saturation in CT and LBM. The normalized $A_{sl}$ is a good indicator of the degree of saturation and can be interpreted as the wetted surface area. In both CT and LBM, the normalized $A_{sl}$, decreases as saturation decreases along primary drainage. During imbibition, $A_{sl}$ increases but does not show any pronounced hysteresis. The wettability of grains is independent of the saturation path.
Although the surface area ($A_{g}$) changes with $S_r$, we do not observe pronounced hysteretic behavior. Unlike the wetted surface area ($A_{sl}$) in~\cref{Fig:interfaces}, LBM simulation shows a pronounced hysteresis in both $A_{sg}$ and $A_{lg}$, with imbibition having a smaller solid-gas interfacial area and a larger liquid-gas interfacial area than drainage at the same saturation. During imbibition, the liquid coating develops at grain surface, separating the gas clusters, causing the gas clusters that adhere to the grains during drainage detach from the grains during imbibition, thereby exposing more surface area to liquid. The increase in the solid-liquid interface is the same as the decrease in the solid-gas interface since the sum of solid-liquid interfacial area ($A_{sl}$) and solid-gas interfacial area ($A_{sg}$) equals the surface area of grains ($A_{solid}$). The sum of solid-gas interfacial area ($A_{sg}$) and liquid-gas interfacial area  ($A_{lg}$) equals the surface area of gas clusters ($A_{g}$). As shown in the LBM section of~\cref{Fig:interfaces}, the points ($A$, $B$, $C$) and ($A'$, $B'$, $C'$) refer to the values of normalized $A_{g}$, $A_{sg}$ and $A_{lg}$ at $S_r=0.5$ during drainage and imbibition ($A_{g}=A_{sg}+A_{lg}$). $A=B+C=0.35+0.09=0.44$, which is similar to $A'=B'+C'=0.3+0.13=0.43$ in imbibition. In other words, the decrease in $A_{sg}$ is the same as the increase in $A_{lg}$, which is why the hysteresis is not observed in the $A_{g}$. 

\subsection{Pore-scale mechanisms governing the hysteresis in WRC}

We explore how the local pore-scale mechanisms control the hysteretic behavior of WRC. To this effect, we compare the number and size distribution of liquid and gas clusters, the morphological change in gas clusters, and the liquid pressure distribution during drainage and imbibition. 

\subsubsection{Number and size distribution of liquid and gas clusters}

\begin{figure}[t!]
  \centering 
  \begin{subfigure}[t]{0.48\columnwidth}
    \centering
    \includegraphics[width=\textwidth]{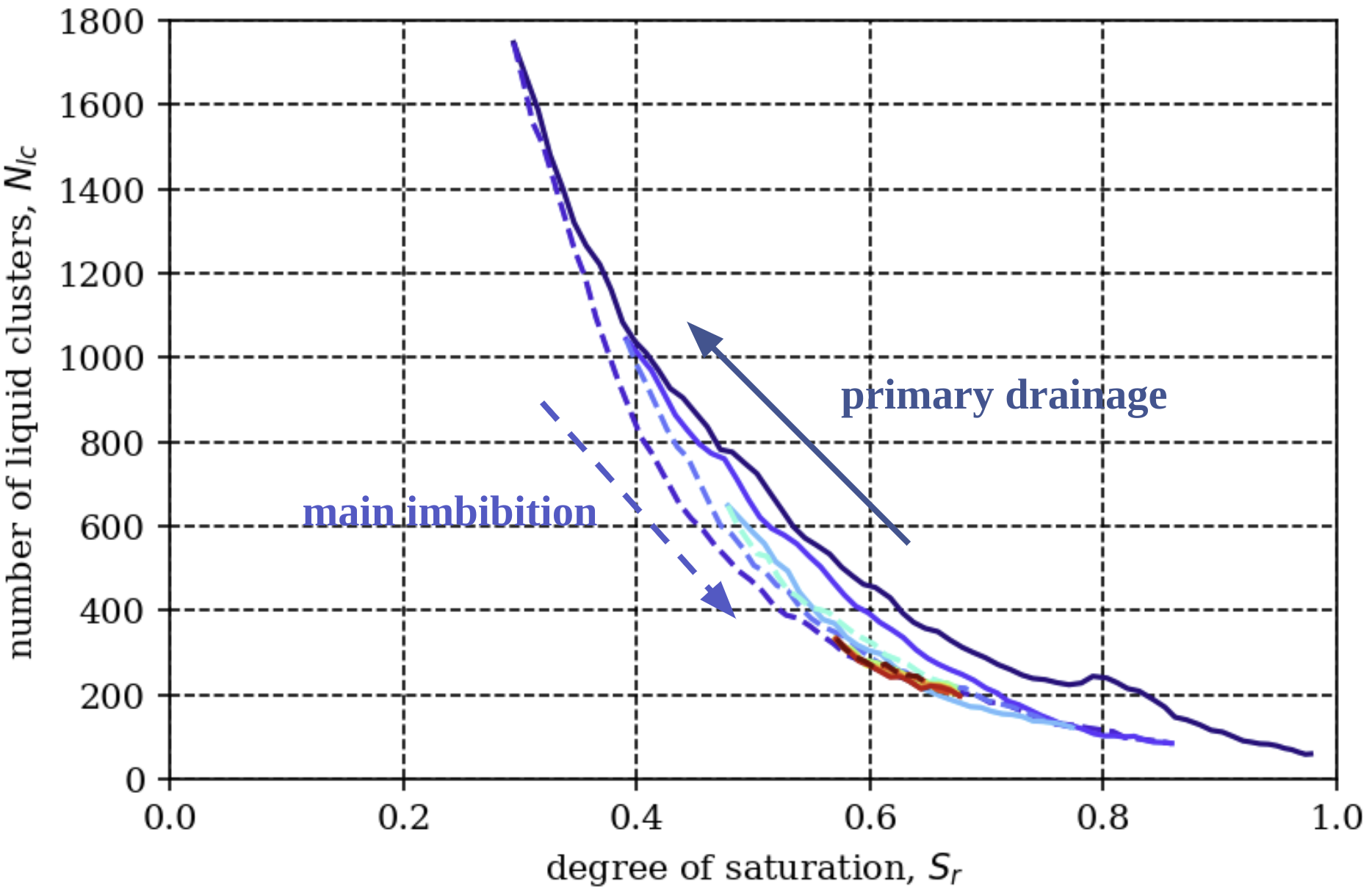}
    \caption{number of liquid clusters}
  \end{subfigure}%
  ~ 
  \begin{subfigure}[t]{0.48\columnwidth}
    \centering
    \includegraphics[width=\textwidth]{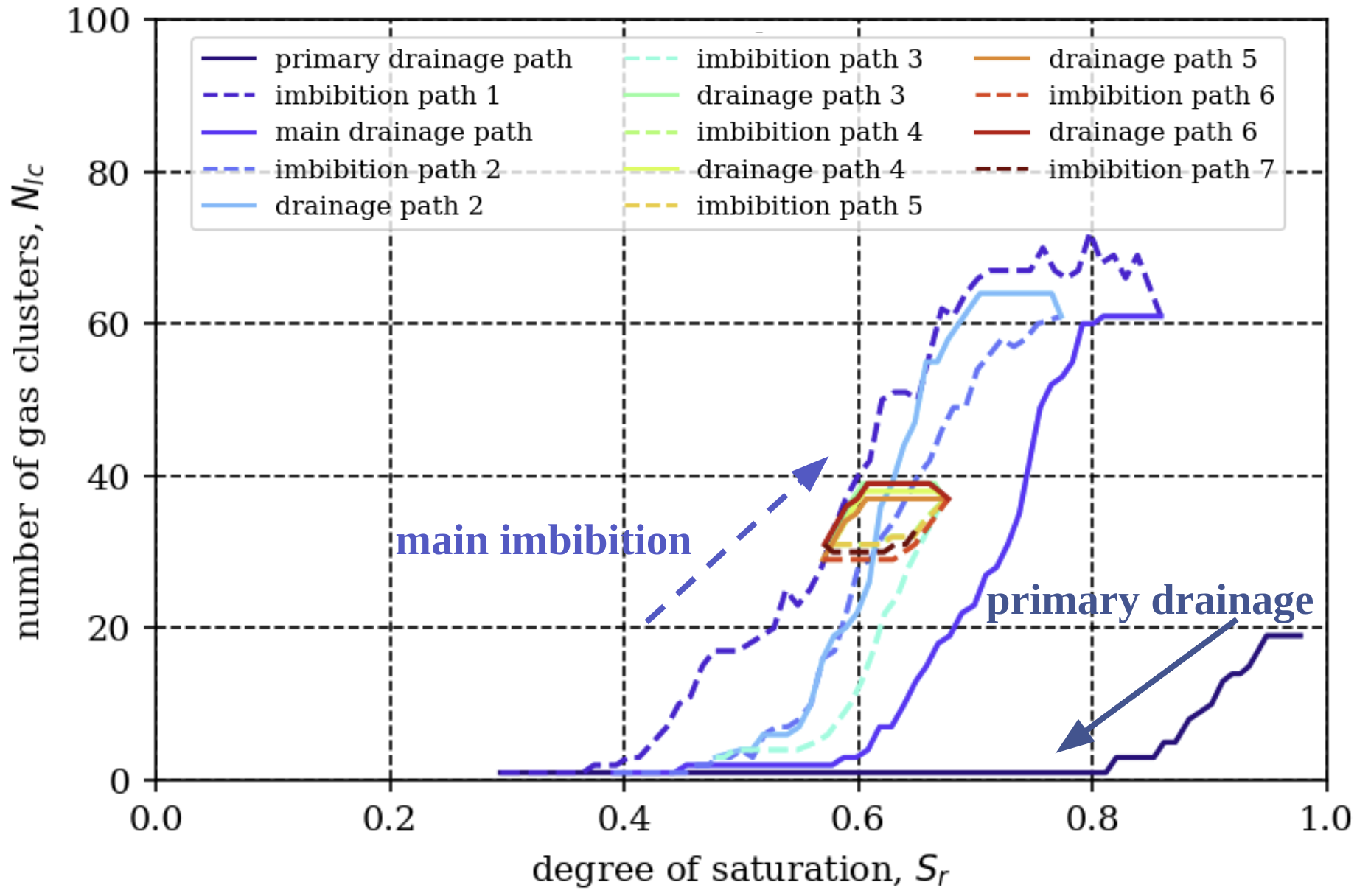}
    \caption{number of gas clusters}
  \end{subfigure}%

\caption{Evolution of (a) the number of liquid clusters and (b) the number of gas clusters \emph{vs.} degree of saturation for Hamburg sand from LBM simulation of cyclic drainage and imbibition. }
\label{Fig:number_of_liq_gas_clusters}
\end{figure}

\Cref{Fig:number_of_liq_gas_clusters} shows the evolution of the number of liquid clusters ($N_{lc}$) and gas clusters ($N_{gc}$) in the LBM simulation. The number of liquid clusters in~\cref{Fig:number_of_liq_gas_clusters}a does not show hysteresis with changes in saturation. In contrast, the number of gas clusters (\cref{Fig:number_of_liq_gas_clusters}b) shows a clear hysteretic behavior with saturation cycles. At the start of primary drainage, we observe 19 gas clusters. Even for a small decrease in saturation (from $S_{r} =0.98$ to $S_{r} =0.8$), the number of gas clusters dramatically decreases, forming a single large gas cluster ($N_{gc}=1$). Only this single gas cluster continues to exist below $S_{r} =0.8$. In other words, after the initial phase separation at the beginning of primary drainage, only the liquid at the liquid-gas interface can transform into gas as drainage continues. As saturation cycle changes to imbibition and $S_r$ increases beyond 0.4, the number of gas clusters begins to increase dramatically, reaching a maximum of 70 distinct gas clusters at $S_{r}=0.8$ as the liquid bridges expands at the grain surface, dividing the largest gas cluster into more gas clusters. 

\begin{figure*}[htbp]
  \centering
  \begin{subfigure}[htbp]{\textwidth}
    \centering
    \includegraphics[width=\textwidth]{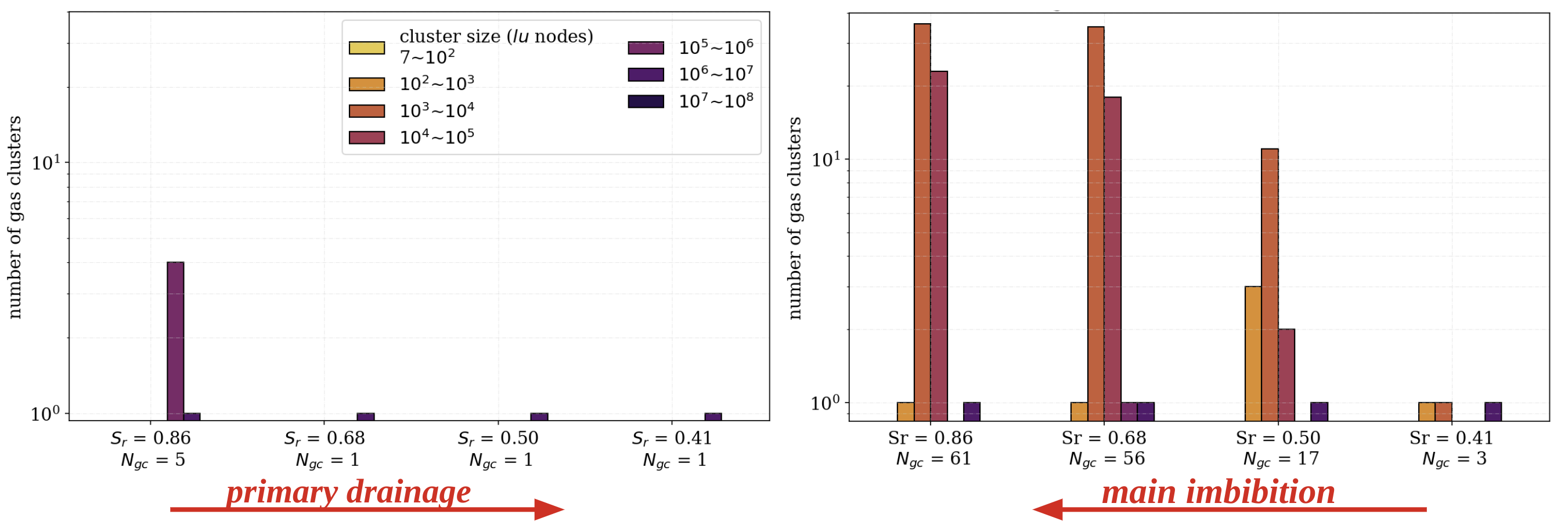}
    \caption{gas cluster size distribution}
  \end{subfigure}%
  
  \begin{subfigure}[htbp]{\textwidth}
    \centering
    \includegraphics[width=\textwidth]{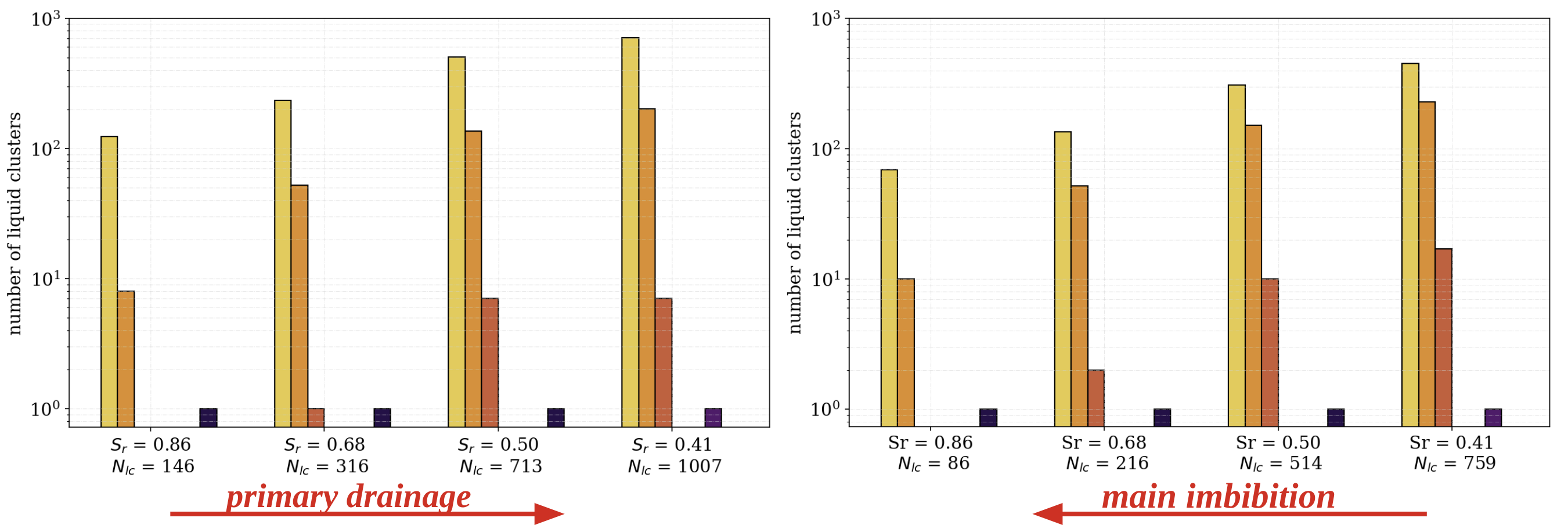}
    \caption{liquid cluster size distribution}
  \end{subfigure}
  \caption{Evolution of liquid and gas cluster size distribution: frequency of liquid and gas cluster size at different constant degrees of saturation during primary drainage and main imbibition based on LBM simulation. }
\label{Fig:liq_gas_clustersize}
\end{figure*}

To see if the size of liquid and gas clusters exhibit hysteretic behavior with saturation cycles, we compute histograms of liquid and gas cluster size distributions during primary drainage and main imbibition.~\cref{Fig:liq_gas_clustersize} shows the evolution of liquid and gas cluster size at different degrees of saturation during primary drainage and main imbibition. 
Although the total volume of liquid and gas clusters is the same at a given saturation, the size distribution of liquid or gas clusters varies. As seen in~\cref{Fig:liq_gas_clustersize}, a single large gas cluster of the size more than $10^6$ nodes persists throughout the primary drainage path. Similarly, a single large liquid cluster of the size more than $10^6$ nodes persist throughout the main imbibition path. We define a large liquid or gas cluster as having more than $10^6$ liquid or gas nodes. We consider a cluster as small if it is smaller than the largest pore which has approximately $10^4$ nodes. Thus, a small cluster has fewer than $10^4$ nodes, a medium cluster has $10^4\sim10^6$ nodes, and a large cluster has more than $10^6$ nodes.

At $S_{r} =0.86$ along the primary drainage (see~\cref{Fig:liq_gas_clustersize}a), only 4 medium gas clusters and 1 large gas cluster exist in the pore space. As saturation decreases, only a single gas cluster exist. In contrast, along imbibition, we see small clusters appear when saturation increases to $S_{r} =0.41$, followed by the growing numbers of medium gas clusters with increasing saturation. During imbibition, as density increases throughout the liquid domain, the liquid bridge first forms between a pair of grains that are closest to each other, or grows within the smallest pore openings. In smaller pore spaces when the liquid bridge grows during imbibition, it separates the largest gas cluster into smaller gas clusters. As imbibition continues, medium gas clusters appear as the expanding liquid bridges transform into liquid clusters that connect multiple grains. Therefore, we observe significant path dependence in both the number and size of gas clusters during drainage and imbibition.

Liquid clusters show no such hysteretic response (see~\cref{Fig:liq_gas_clustersize}a). A single large liquid cluster coexists with small and medium-size liquid clusters throughout the primary drainage and main imbibition. The proportion of small liquid clusters decreases as saturation increases, and vice versa.

\begin{figure*}[htbp]
\begin{center}
\includegraphics[width=0.65\columnwidth]{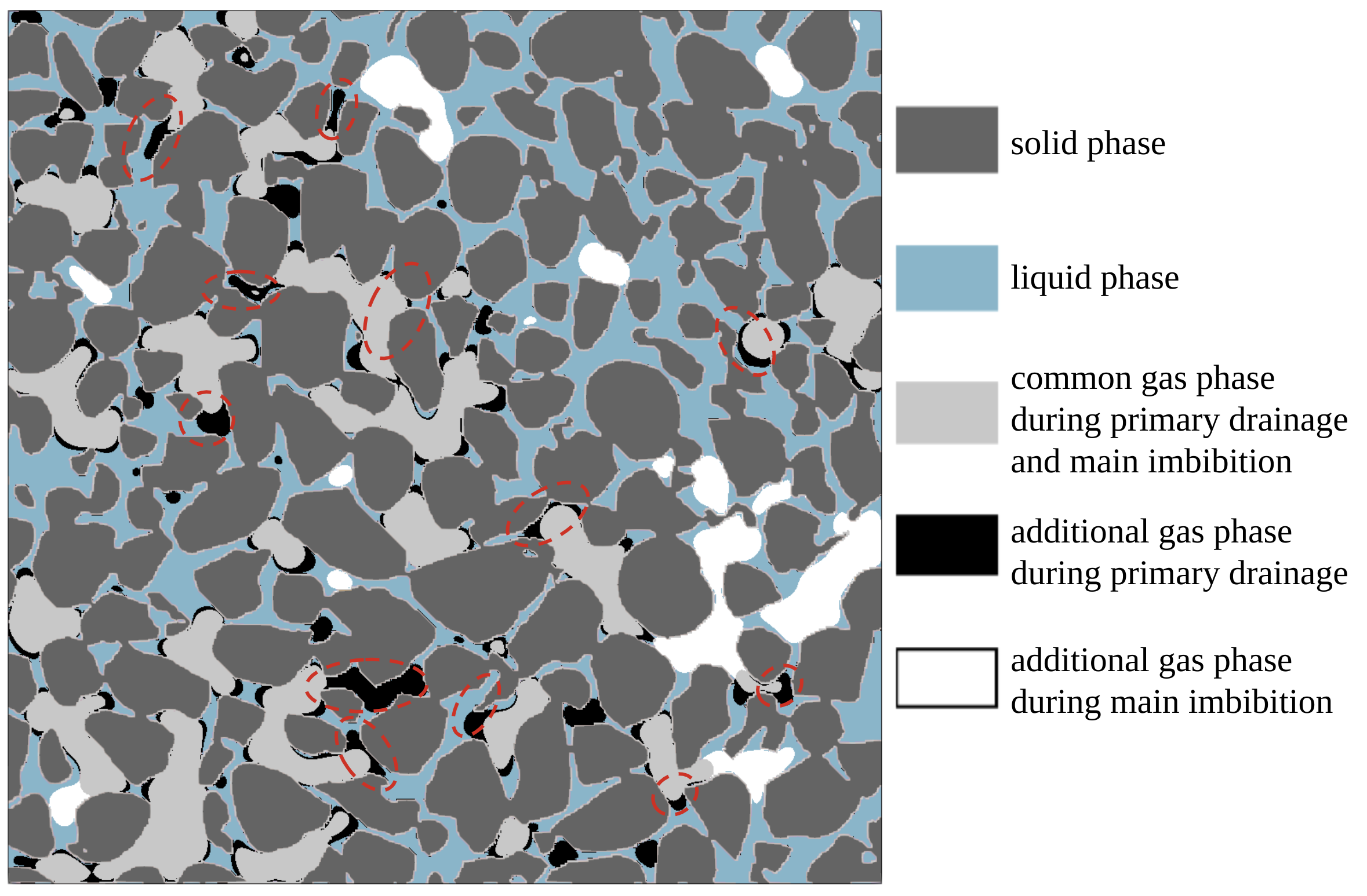}
\end{center}
\caption{Overlapping 2D slices of liquid and gas phases at mid-height of the sample at $S_{r} = 0.59$ along the primary drainage and main imbibition; the colors of liquid and gas phases are marked in the legend; the dashed red ovals partially highlight the additional gas phase that intrudes deeper into the pores during the primary drainage.}
\label{Fig:Morphological_gas_clusters}
\end{figure*}

\subsubsection{Morphological transformation of gas clusters}
This section investigate how the gas cluster morphology evolves with the change in saturation at the pore scale. In order to understand the differences in the distribution of the fluid phase during drainage and imbibition, we overlay the fluid phase distributions and highlight the similarities and differences between the drainage and imbibition in~\cref{Fig:Morphological_gas_clusters}. The gray area in~\cref{Fig:Morphological_gas_clusters} shows the common gas phase between drainage and imbibition. The black and white regions show the differences in distribution; the black region represents the additional gas phase during primary drainage and the white zones denote the additional gas phase in main imbibition. Observing these additional phases shows that in primary drainage, the gas clusters localize around grain surfaces (selected regions are highlighted as dashed red ovals in~\cref{Fig:Morphological_gas_clusters}). These gas clusters take the shape of the grain surface to which they are connected, resulting in irregular shapes. These irregular shapes lead to a small overall radius of curvature at the liquid-gas interface thus generating larger suction during drainage. During imbibition, the additional gas clusters (shown in white) tend to reside near the center of the pore spaces. As the liquid expands, the gas clusters retreat to the center of the pores taking a more spherical shape. The overall radius of curvature at the liquid-gas interface is larger, leading to a smaller suction during imbibition.~\citet{hosseini2022investigating} made a similar observation for the behavior of gas clusters for a sample made of spherical grains. The expansion of gas clusters during pore emptying is restrained by the size of the pore openings around them, whereas the shrinkage of gas clusters during pore filling is independent of the size of the pore openings.

\begin{figure*}[htbp]
  \centering
  \begin{subfigure}[htbp]{\textwidth}
    \centering
    \includegraphics[width=\textwidth]{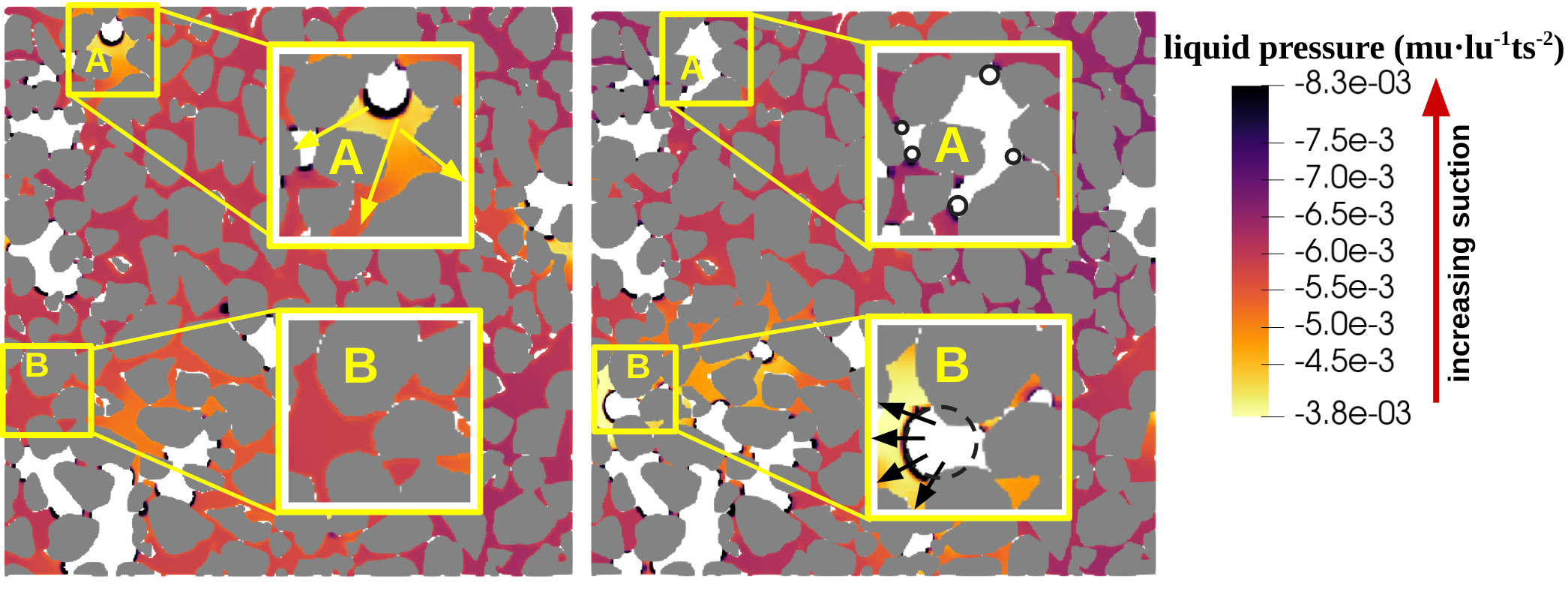}
    \caption{liquid pressure distribution at $S_{r} = 0.86$ (left) and $S_{r} = 0.83$ (right)}
  \end{subfigure}%
  
  \begin{subfigure}[htbp]{\textwidth}
    \centering
    \includegraphics[width=\textwidth]{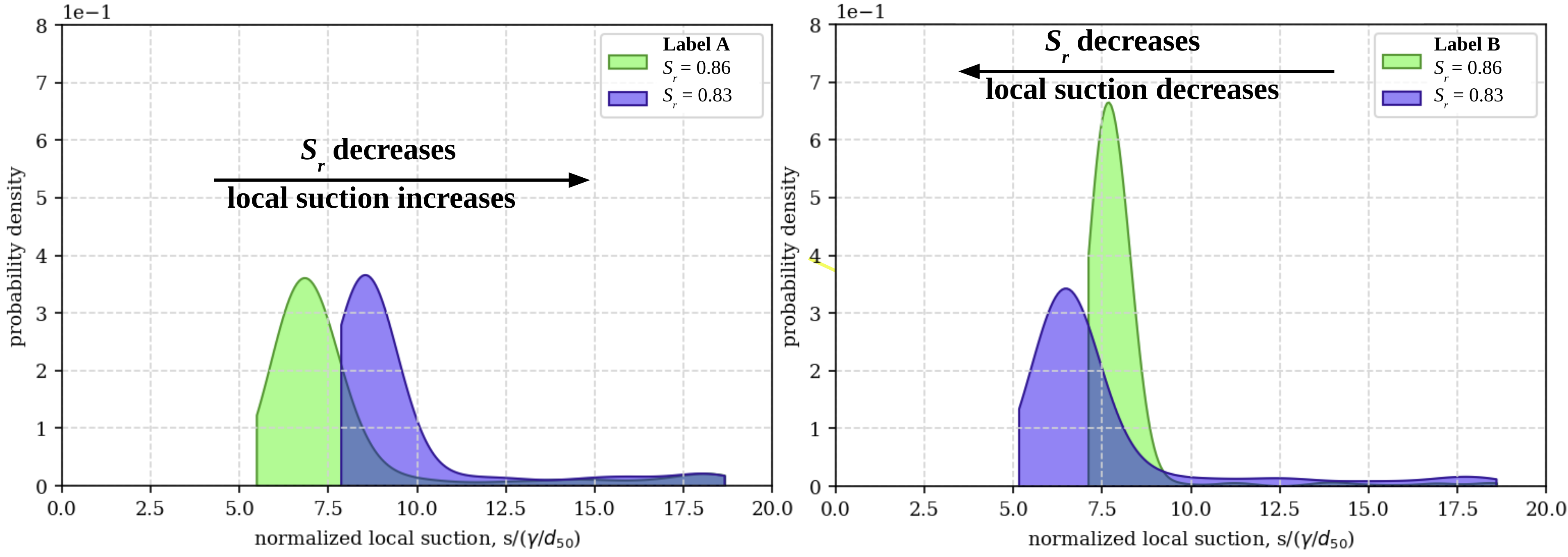}
    \caption{probability density of local suction in Label A and B}
  \end{subfigure}
  \caption{(a) Liquid pressure distribution and (b) probability density of local suction  in labelled local zones in yellow boxes at mid-height of the sample at $S_{r} = 0.86$ and $S_{r} = 0.83$ along primary drainage; the solid and dashed black circles represent the radii of curvature at the interface between liquid and gas. }
\label{Fig:Liquid_pressure_distribution_drain}
\end{figure*}

\begin{figure*}[htbp]
  \centering
  \begin{subfigure}[htbp]{\textwidth}
    \centering
    \includegraphics[width=\textwidth]{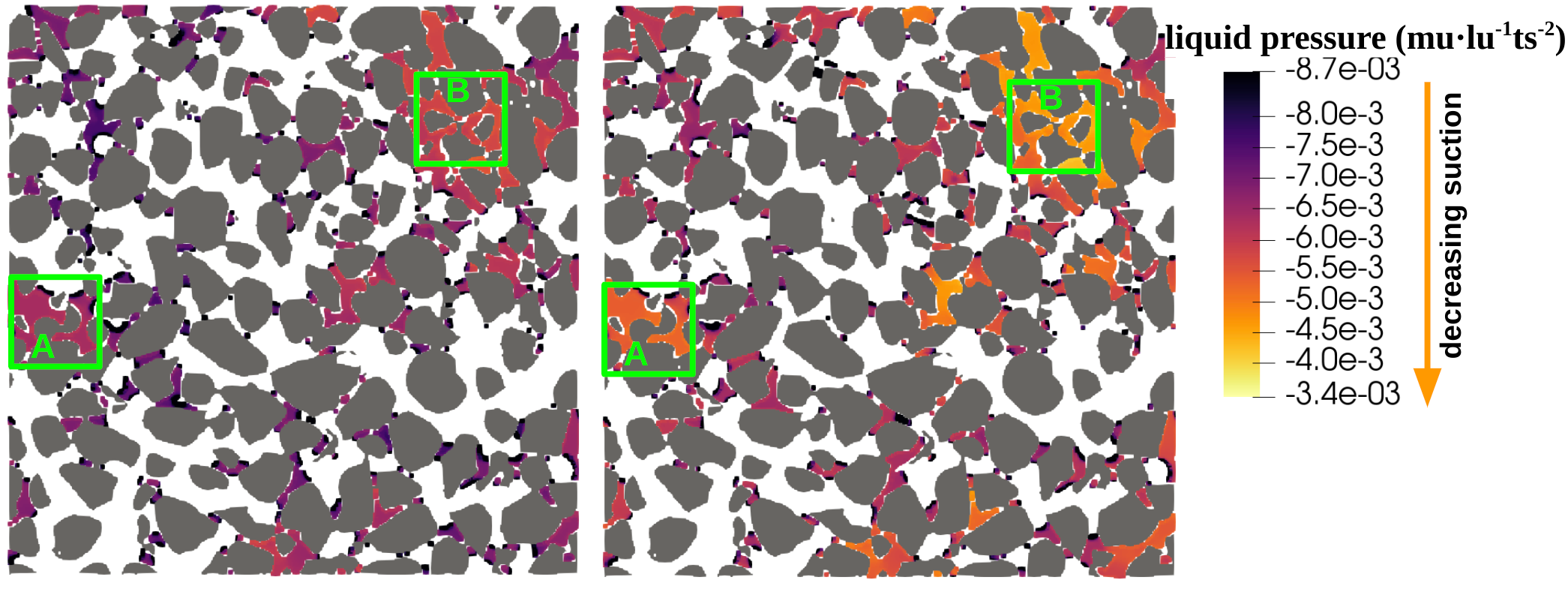}
    \caption{liquid pressure distribution at $S_{r} = 0.35$ (left) and $S_{r} = 0.38$ (right)}
  \end{subfigure}%
  
  \begin{subfigure}[htbp]{\textwidth}
    \centering
    \includegraphics[width=\textwidth]{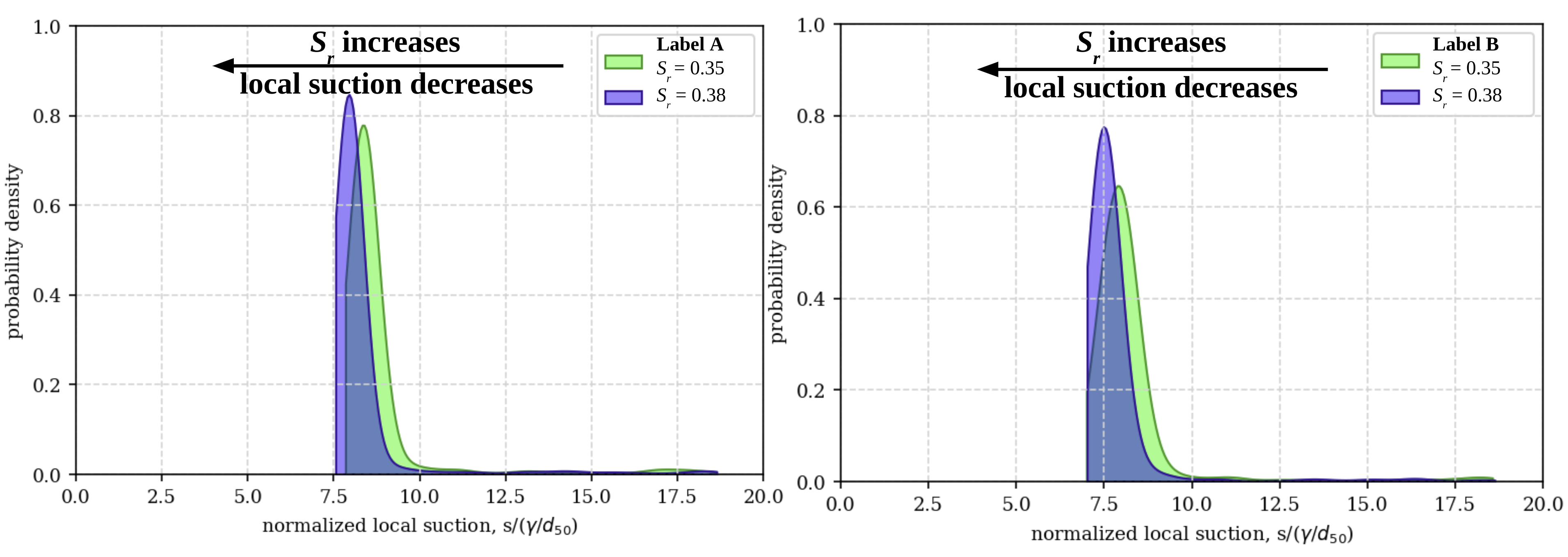}
    \caption{probability density of local suction in Label A and B}
  \end{subfigure}
  \caption{(a) Liquid pressure distribution and (b) probability density of local suction in labelled local zones in green boxes at mid-height of the sample at $S_{r} = 0.35$ and $S_{r} = 0.38$ along main imbibition. }
\label{Fig:Liquid_pressure_distribution_imb}
\end{figure*}

\subsubsection{Liquid pressure distribution along drainage and imbibition paths}

The morphology of gas clusters varies between drainage and imbibition. We now examine how local suction evolves as gas clusters expand into or retreat from small pore spaces, over a small saturation range during drainage or imbibition. We identify two classes of pore space: one where the gas cluster is entering smaller pore throats (Label A in~\cref{Fig:Liquid_pressure_distribution_drain}a) and another where the gas clusters are entering larger pore throats (Label B in~\cref{Fig:Liquid_pressure_distribution_drain}a). We calculate the local suction as the difference between gas pressure and liquid pressure in the pore throat. Since gas pressure is almost constant throughout the simulation, the local suction is proportional to the liquid pressure. As the global saturation decreases by a small amount from $S_{r}=0.86$ to $S_{r}=0.83$, the gas cluster expands into the nearby smaller pore throats, the radius of curvature decreases (see the small radii of circles marked in Label A of~\cref{Fig:Liquid_pressure_distribution_drain}a.) ~\Cref{Fig:Liquid_pressure_distribution_drain}b shows the probability density of the local suction at all fluid nodes inside the Label A and B in~\Cref{Fig:Liquid_pressure_distribution_drain}a. According to the Young-Laplace-equation (\Cref{eq:Young-Laplace}), a decrease in the radius of curvature causes an increase in local suction for pore label A (\cref{Fig:Liquid_pressure_distribution_drain}b). In contrast, as the gas clusters expand into larger pore throats as saturation decreases, the radius of curvature increases (see the large circle marked in Label B of~\cref{Fig:Liquid_pressure_distribution_drain}a), causing a drop in the local suction, shifting the curves in~\cref{Fig:Liquid_pressure_distribution_drain}b to the left. Overall, as the saturation decreases, depending on the pore size in which the gas clusters are located, we observe an increase or a decrease in local suction values. The fluctuations in the overall suction along the primary drainage path (see~\cref{Fig:comparison_of_wrcs}) are a result of averaging over these local variations.

\Cref{Fig:Liquid_pressure_distribution_imb} shows the evolution of liquid pressure distribution in the sample as the global saturation increases from $S_{r} = 0.35$ to $S_{r} = 0.38$ along the main imbibition. Similar to our analysis in primary drainage, we select two classes of pore spaces: one where the liquid cluster is surrounded by gas content (Label A in~\cref{Fig:Liquid_pressure_distribution_imb}a) and another where the liquid cluster is enclosed by grains (Label B in~\cref{Fig:Liquid_pressure_distribution_imb}a). In contrast to the behaviors of gas clusters during drainage, the local liquid pressure in both cases decreases. This behavior is observed in all pore spaces, where a uniform decrease in suction throughout the liquid zone results in a smooth imbibition path as shown in~\cref{Fig:comparison_of_wrcs}b. Based on the observations above, we can summarize that during drainage but not during imbibition, the local suction response is dependent on pore size.

\subsection{Pore size dependency of water retention behavior}
\label{sec:Pore size dependency}

We use the ``sphere placement'' (chamber size) method to classify pore sizes (see~\cref{sec:material_and_methods}) and analyze the influence of pore size on the hysteretic behavior~\citep{sweeney2003pore}.
In the ~\SI{8}{\milli\meter\cubed} sample, we identify a log-normal distribution of 63,713 pores, of which 38,867 have a pore radius smaller than 3 lu. Small pores predominate the pore space; however, pore with radii greater than 5 lu occupy more than 85\% of the pore space. We now evaluate how local suction varies based on the pore size.

\begin{figure*}[t!]
  \centering
   \begin{subfigure}[t]{0.49\columnwidth}
    \centering
    \includegraphics[width=\columnwidth]{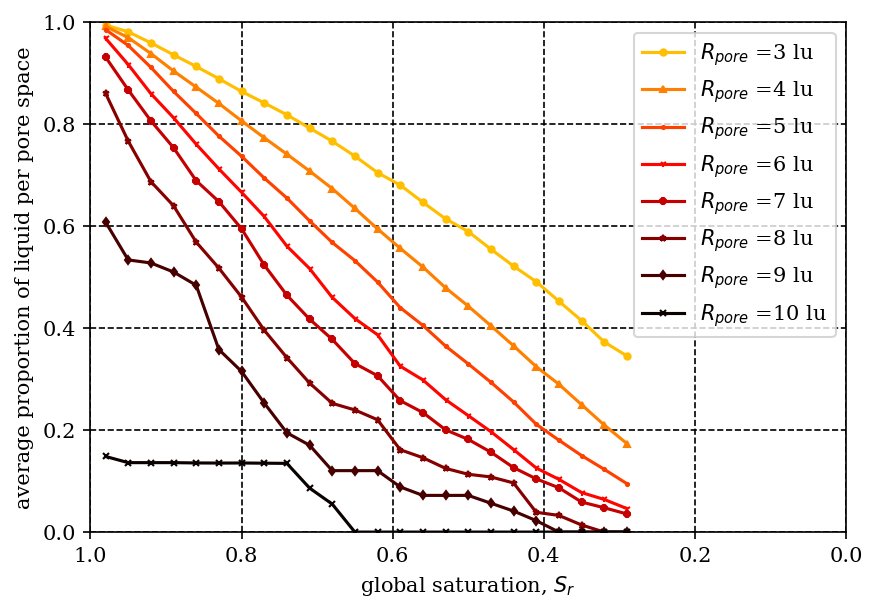}
    \caption{average proportion of liquid during primary drainage}
  \end{subfigure}%
  ~ 
  \begin{subfigure}[t]{0.49\columnwidth}
    \centering
    \includegraphics[width=\columnwidth]{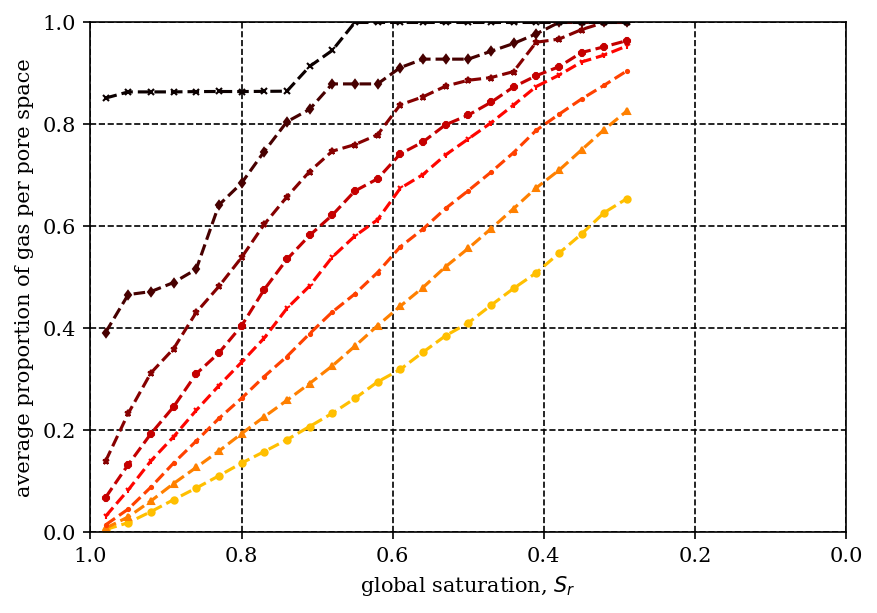}
    \caption{average proportion of gas during primary drainage}
  \end{subfigure}
  
     \begin{subfigure}[t]{0.49\columnwidth}
    \centering
    \includegraphics[width=\columnwidth]{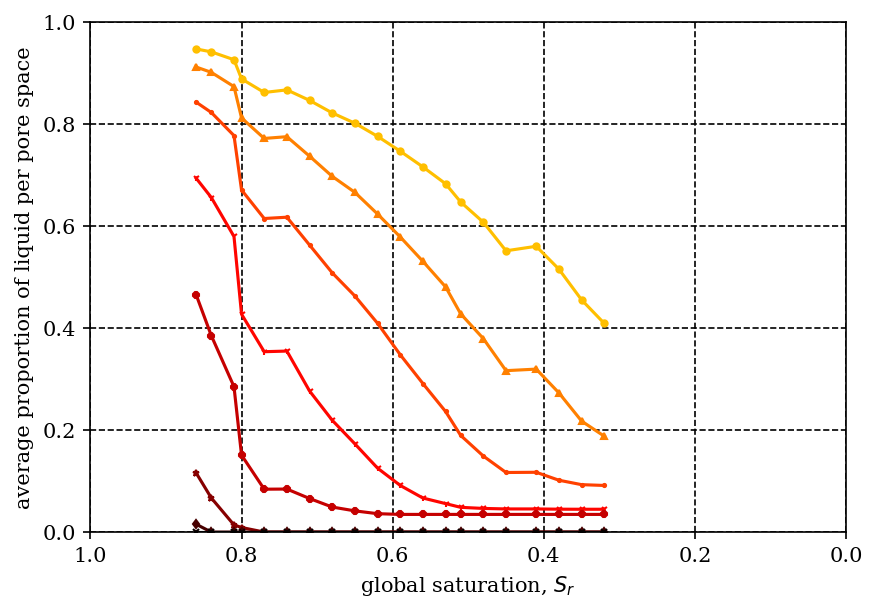}
    \caption{average proportion of liquid during main imbibition}
  \end{subfigure}%
  ~ 
  \begin{subfigure}[t]{0.49\columnwidth}
    \centering
    \includegraphics[width=\columnwidth]{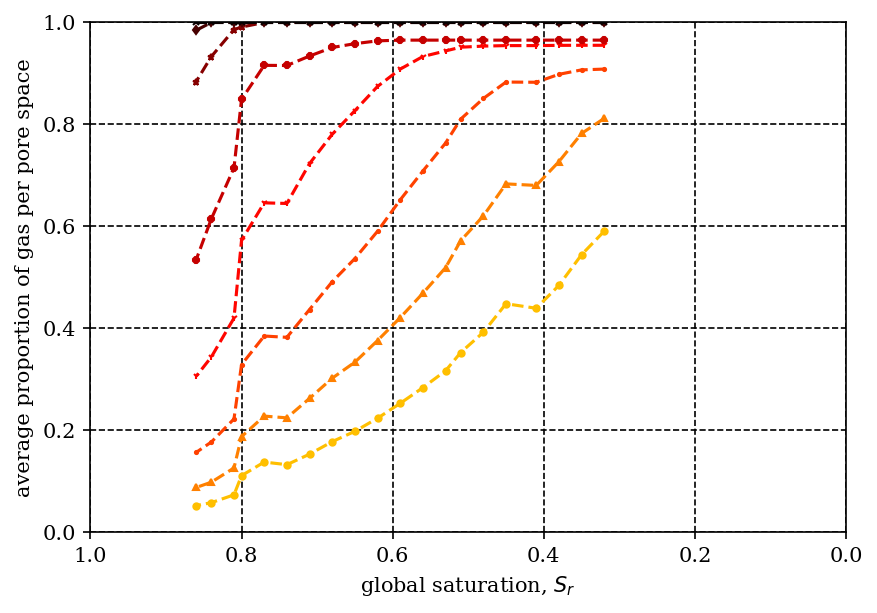}
    \caption{average proportion of gas during main imbibition}
  \end{subfigure}
  \caption{Proportion of liquid and gas per pore space for different pore radii (3 lu$<=R_{pore} <=$10 lu) during primary drainage and relative liquid nodes during main imbibition; pores with a pore radius greater than $10$ lu are not chosen because there are only one or two such pores in the entire pore space}
\label{Fig:gasorliquid_per_pore}
\end{figure*}

\begin{figure}[t!]
  \centering
  \begin{subfigure}[t]{0.48\columnwidth}
    \centering
    \adjincludegraphics[width=\columnwidth,trim={0 {0.08\width} 0 0},clip]{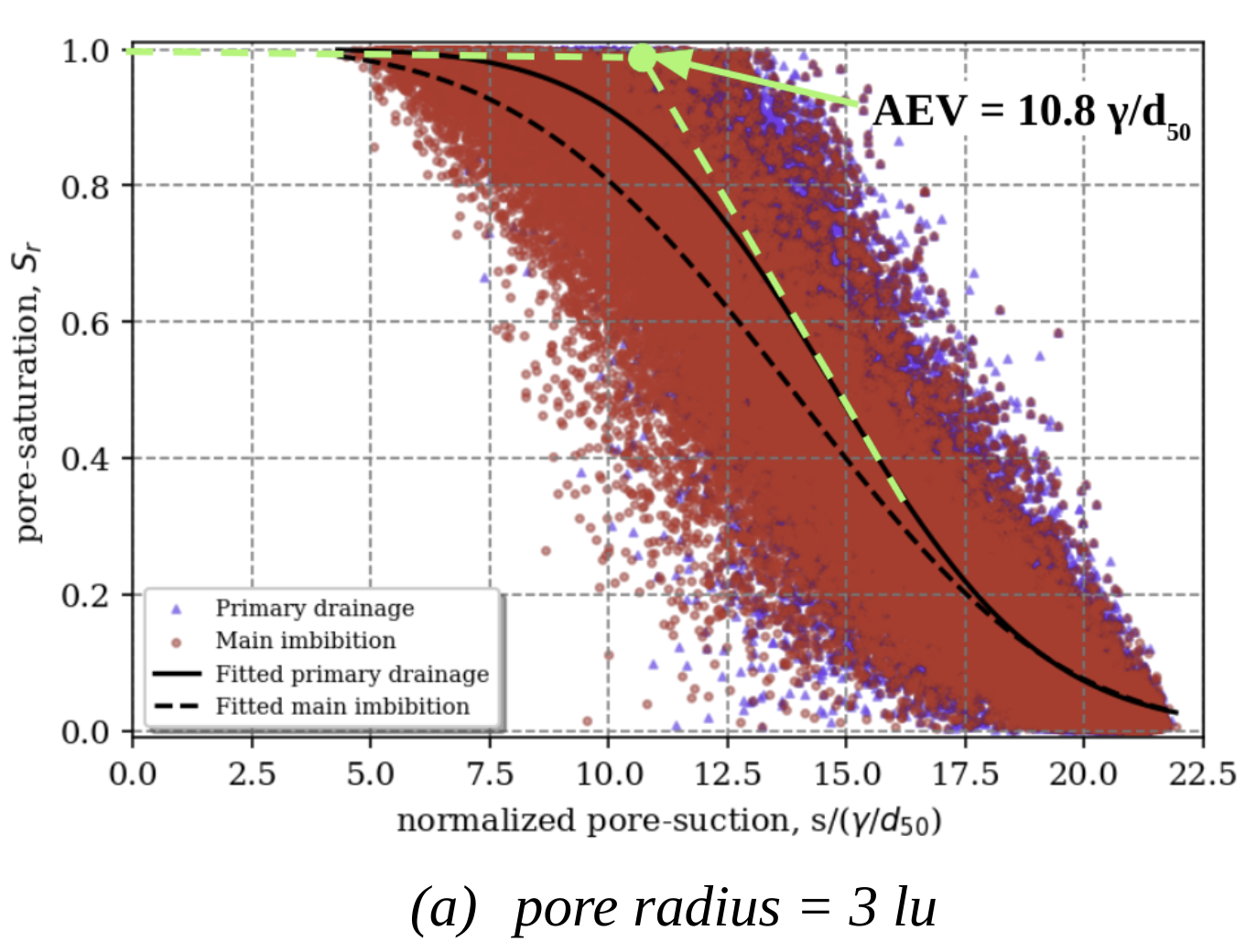}
    \caption{$R_{pore}$ = 3 lu}
  \end{subfigure}%
  ~ 
  \begin{subfigure}[t]{0.48\columnwidth}
    \centering
    \adjincludegraphics[width=\columnwidth,trim={0 {0.08\width} 0 0},clip]{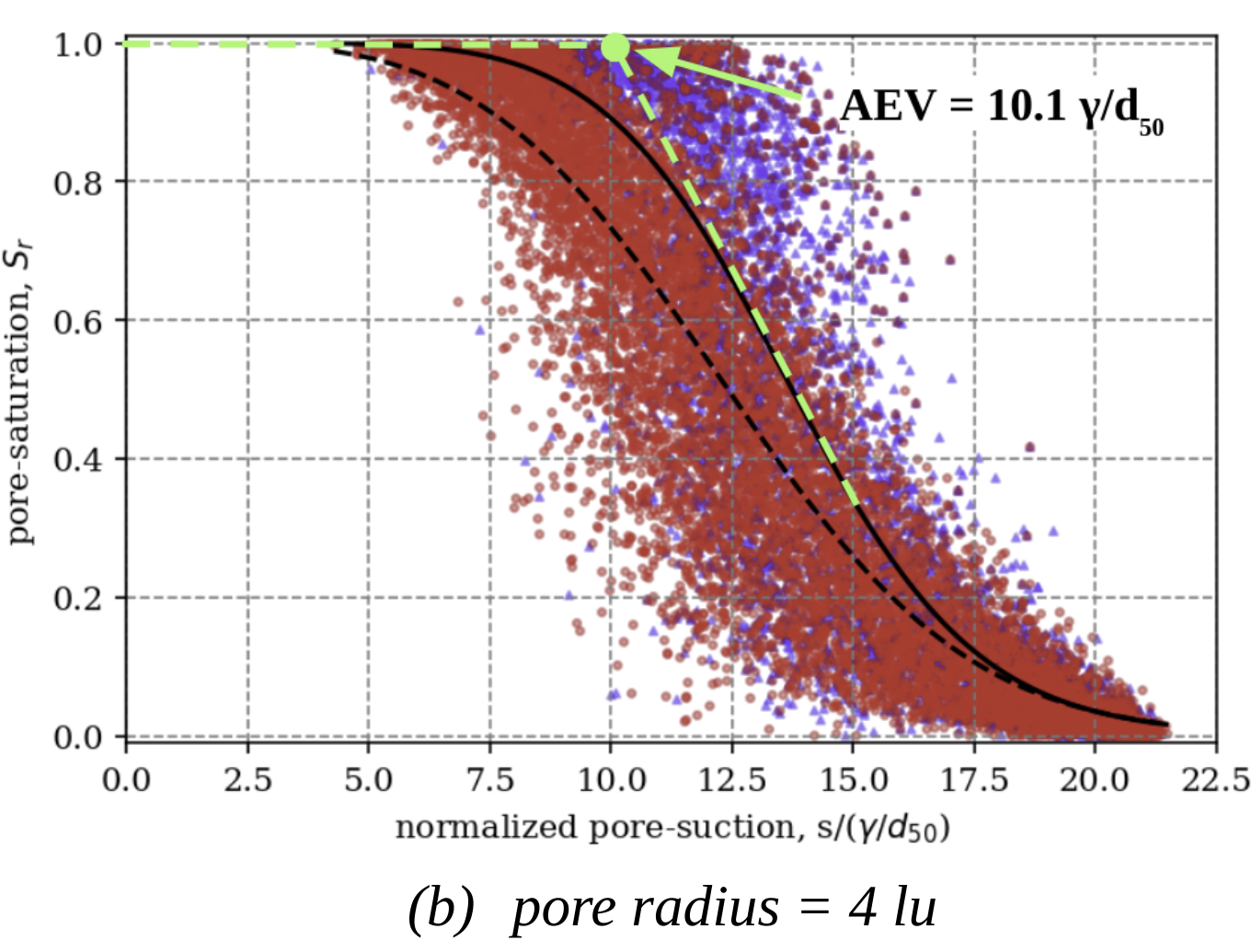}
    \caption{$R_{pore}$ = 4 lu}
  \end{subfigure}
  
  \begin{subfigure}[t]{0.48\columnwidth}
    \centering
    \adjincludegraphics[width=\columnwidth,trim={0 {0.08\width} 0 0},clip]{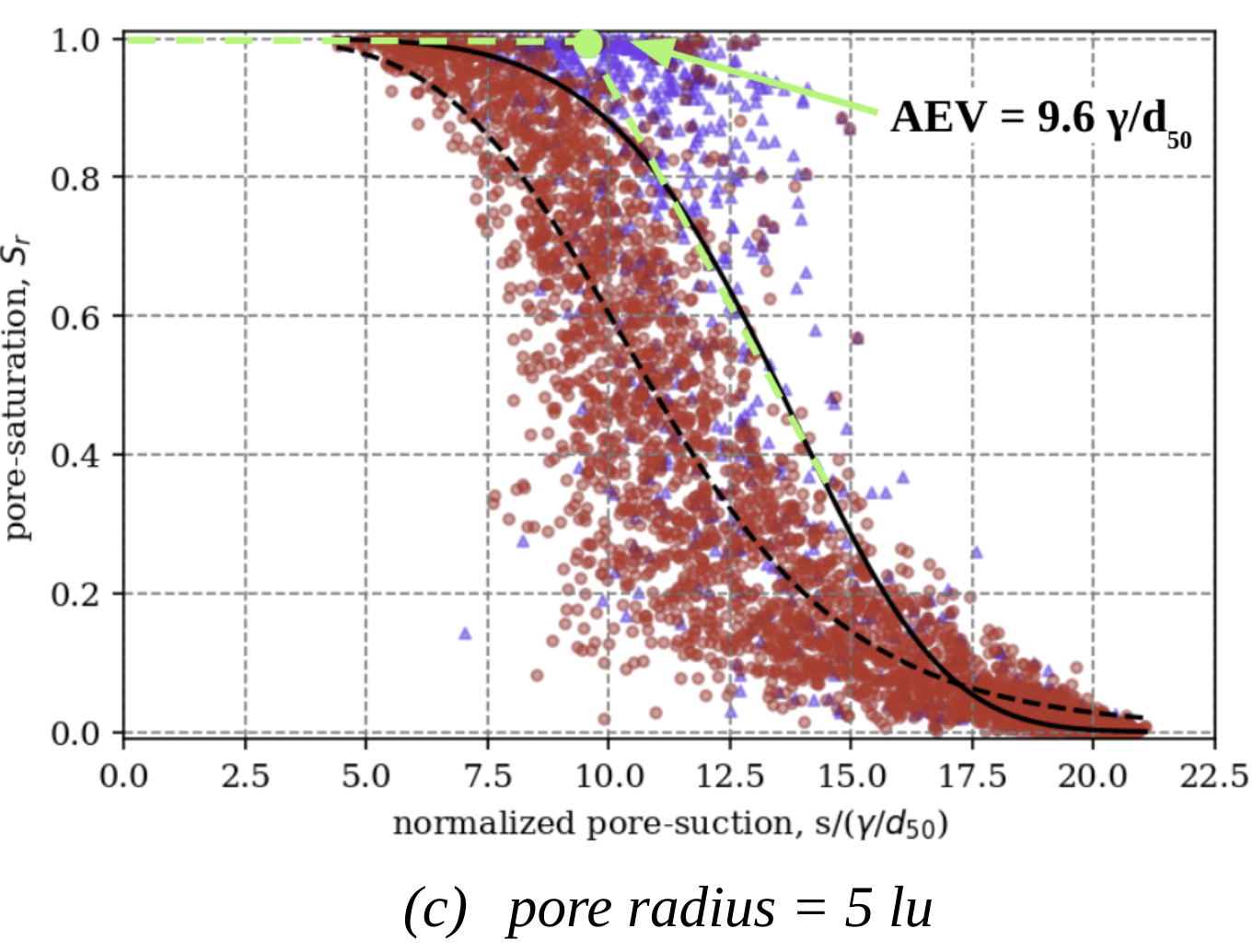}
    \caption{$R_{pore}$ = 5 lu}
  \end{subfigure}%
  ~ 
  \begin{subfigure}[t]{0.48\columnwidth}
    \centering
    \adjincludegraphics[width=\columnwidth,trim={0 {0.08\width} 0 0},clip]{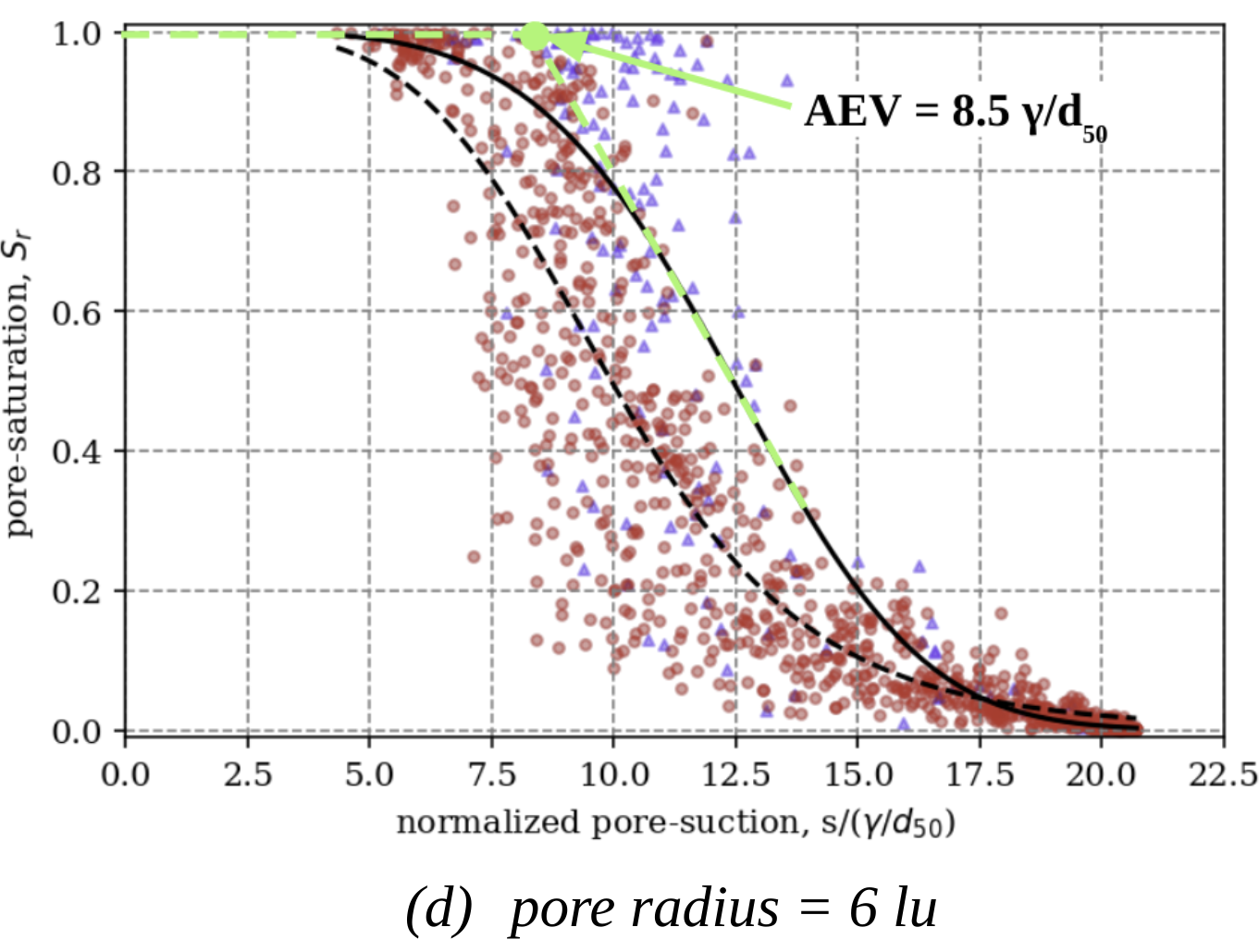}
    \caption{$R_{pore}$ = 6 lu}
  \end{subfigure}
  
   \begin{subfigure}[t]{0.48\columnwidth}
    \centering
    \adjincludegraphics[width=\columnwidth,trim={0 {0.08\width} 0 0},clip]{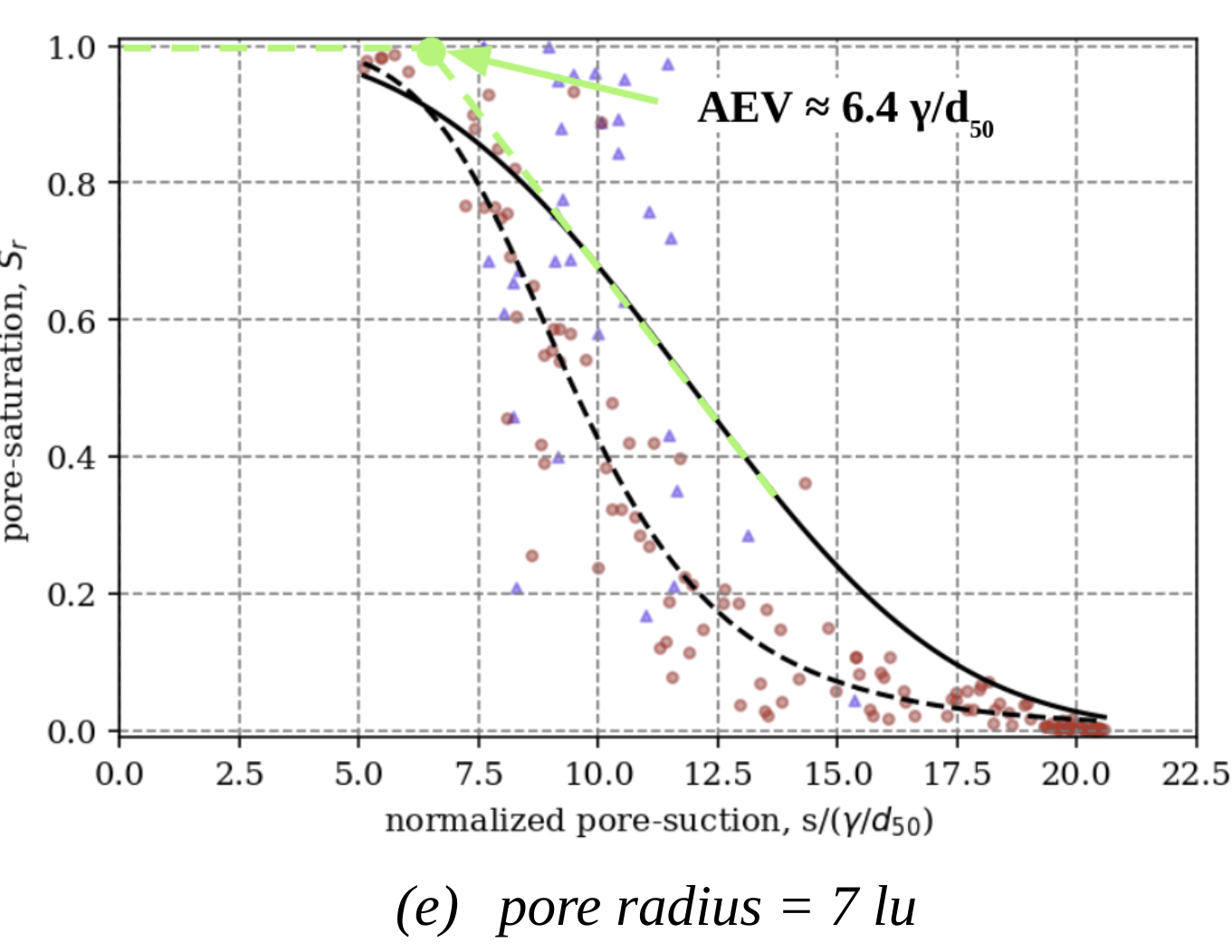}
    \caption{$R_{pore}$ = 7 lu}
  \end{subfigure}%
  ~ 
  \begin{subfigure}[t]{0.48\columnwidth}
    \centering
    \adjincludegraphics[width=\columnwidth,trim={0 0 0 0},clip]{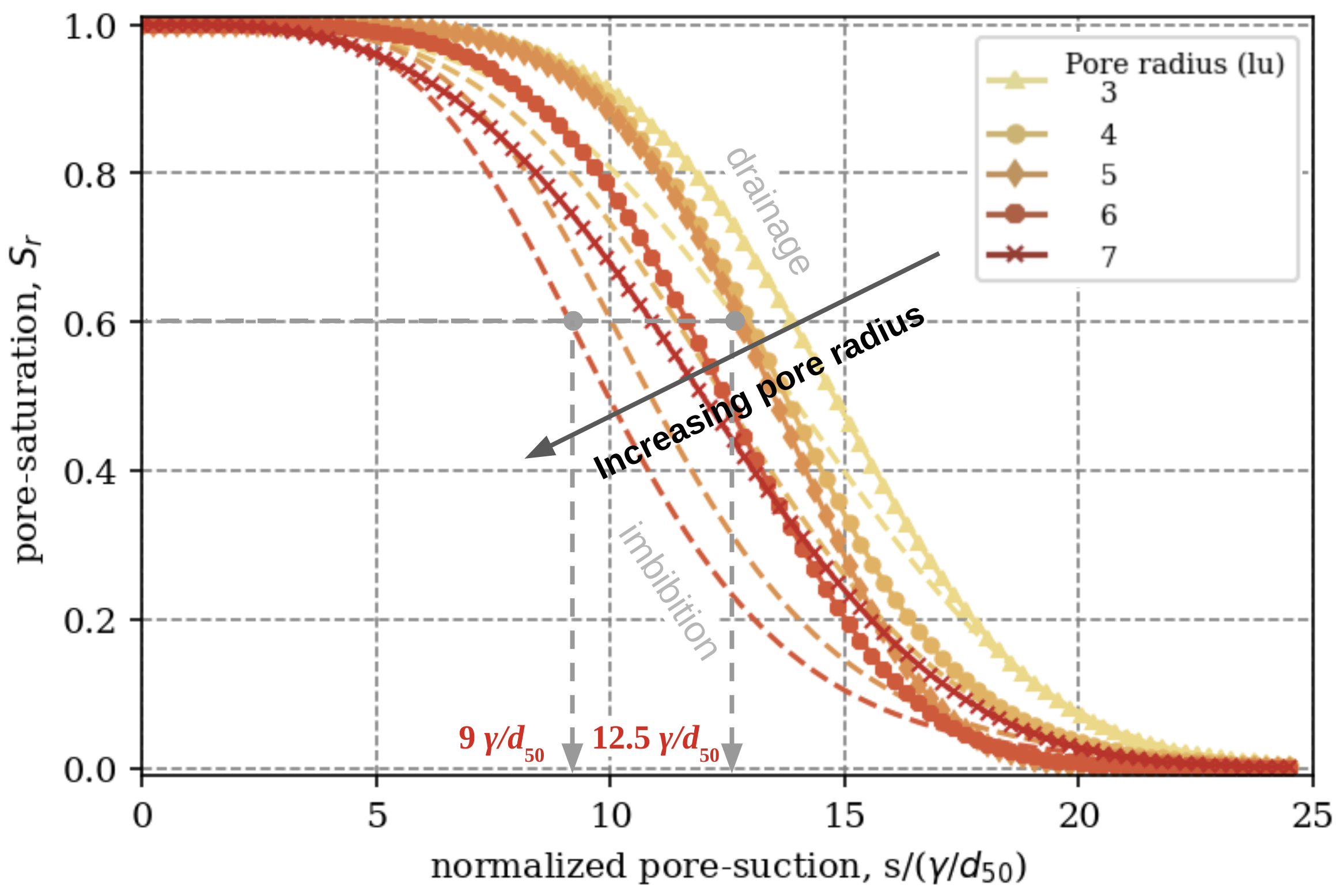}
    \caption{fitted WRCs for increasing pore radii}
  \end{subfigure}
  \caption{Pore saturation versus pore suction for different pore sizes during primary drainage and main imbibition based on the LBM simulation of the hydraulic CT experiment on Hamburg sand.}
\label{Fig:fitted_pore_wrcs}
\end{figure}

\begin{figure*}[htbp]
\begin{center}
\includegraphics[width=0.6\columnwidth]{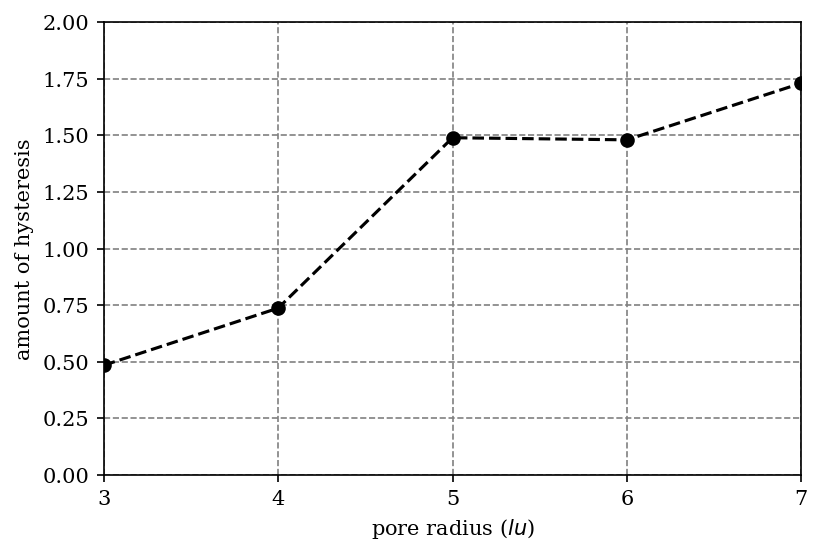} 
\caption{Amount of hysteresis versus pore radius based on the LBM simulation of the hydraulic CT experiment on Hamburg sand. The amount of hysteresis is defined as the area between the primary drainage and main imbibition paths of the pore-based WRCs as shown in \cref{Fig:fitted_pore_wrcs} for $0.1< S_{r}< 0.85$.}
\label{Fig:hysteresis_vs_pore_radii}
\end{center}
\end{figure*}

To reveal the path-dependence of pore emptying and pore filling, we compute the proportion of liquid and gas per pore space for a given pore size along primary drainage and main imbibition paths, as shown in~\cref{Fig:gasorliquid_per_pore}. During drainage (\cref{Fig:gasorliquid_per_pore}a-b), the proportion of gas increases and the proportion of liquid decreases at the same rate. All the pores drain simultaneously regardless of pore size. The slope of both liquid and gas proportion with respect to $S_r$ increases with increasing pore size, which means large pores drain faster than small pores. When saturation decreases to $S_r = 0.67$, the largest pores ($R_{pore} =10\,$lu) are emptied as the proportion of gas reaches 1 and proportion of liquid reaches 0. 

During imbibition (\cref{Fig:gasorliquid_per_pore}c-d), pores fill according to the pore sizes, \ie pores of $R_{pore} < 5\,$lu immediately begins saturating as imbibition begins at $S_r =0.3$, then pores of $R_{pore}=5\,$lu start saturating at $S_r = 0.36$, followed by pores of $R_{pore}=6\,$lu at $S_r=0.5$; large pores of $R_{pore}=7\,$lu start saturating only as global saturation reaches $S_r=0.65$, while no pore filling occurs in larger pores ($R_{pore} >= 8\,$lu) throughout the main imbibition.
We evaluate how the local WRC varies with pore size, which range from 3 to 7 lu, as shown in~\cref{Fig:fitted_pore_wrcs}. We also plot the fitted van Genuchten model to the data~\citep{van1980closed}, which is given as:
\begin{equation}
S_{r}=(\frac{1}{1+(\alpha \cdot s)^{n}})^{m},
\end{equation}
where $S_{r}$ and $s$ are the degree of saturation and the suction at the local pore space, $\alpha$, $n$, and $m$ are the fitting parameters.
\cref{Fig:fitted_pore_wrcs} shows the pore-based WRCs are also hysteretic, exhibiting greater hysteresis with increase in pore radius. A large pore allows a larger radius of curvature at the capillary meniscus, resulting in a lower suction, as previously denoted in~\cref{Fig:Liquid_pressure_distribution_drain} labels A and B. Thus, as pore size increases, the pore suction decreases at the same pore saturation along both drainage and imbibition (see~\cref{Fig:fitted_pore_wrcs}f). For example, at the pore saturation of $S_r=0.6$, the pore suction decreases from 12.5 $\gamma/d_{50}$ for a pore size of $R_{pore} = 3\,$lu to 9 $\gamma/d_{50}$ for a pore size of $R_{pore} = 7\,$lu during imbibition. Accordingly, AEV along the drainage path also decreases as pore size increases. 

To quantify the hysteresis of pore-based WRCs, we compute the area between the primary drainage and main imbibition paths of the pore-based WRCs over a fixed range of saturation ($0.3< S_{r}< 0.85$). We select this range of saturation because it falls within both primary drainage and main imbibition paths in the experiment.~\cref{Fig:hysteresis_vs_pore_radii} summarizes the amount of hysteresis for different pore radii from 3 to 7 lu. As pore size increases, the amount of hysteresis also increases.
The increased hysteresis mainly results from the decrease in pore suction during imbibition in larger pores (see~\cref{Fig:fitted_pore_wrcs}f). During imbibition, small pores fill up first, whereas large pores are unsaturated and have capillary menisci. As saturation increases, the pore suction for large pores continues to decrease until the large pores are completely filled. At the same pore saturation, large pores have lower local suction than small pores. Hence, large pores exhibit a stronger hysteresis than smaller pores.

\section{Summary}
\label{sec:summary}
We use the multiphase LBM to investigate the water retention behavior of granular soils using in situ CT experiment with cyclic drainage and imbibition paths. The multiphase LBM reproduces the hydraulic path dependence, the AEV value of WRC, as well as oscillation along drainage paths observed in the CT experiment. To examine the hysteresis at the pore-scale, we compare the spatial distribution of liquid and gas phases between CT and LBM.
LBM captures the transition between successive liquid statistics along the primary drainage and main imbibition paths. 
During primary drainage, only the liquid at the existing liquid-gas interface can transform into gas. In contrast, during main imbibition, the gas can transform into liquid at solid-liquid interface or within liquid. Therefore, we observe pronounced hysteresis in gas-related statistics, \ie the liquid-gas interfacial area, number and size distribution of gas clusters. In contrast, the liquid-related statistics, \ie the liquid-solid interface and number and size distribution of liquid clusters, show no or minor hysteresis and indicates grain wettability is independent of the drainage path.
We examine the morphological changes in gas content at the pore-scale. During drainage, gas clusters are irregular in shape and localized at the grain surface (see~\cref{Fig:Morphological_gas_clusters}), which leads to a smaller overall radius of curvature at the capillary menisci and a larger suction than imbibition. Also, over a small saturation range, the local suction increases in pore spaces where gas clusters enter through small pore openings while the local suction decreases in pore spaces where gas clusters enter through large pore openings. Therefore, the morphology of gas clusters governs the overall suction response, including hysteresis and oscillations along drainage. In contrast, local suction during imbibition decreases uniformly independently of the characteristics of pore spaces. 

We reveal a pore size dependent hysteretic behavior of WRC. Larger pores are emptied sooner but filled later than the smaller pores. Larger the pores show a lower the suction as gas clusters inside large pores can sustain a larger radius of curvature. Furthermore, large pores exhibit greater hysteresis in local suction than small pores as their pore suction keeps decreasing after small pores are fully saturated during imbibition.

\nolinenumbers

\begin{acknowledgements}

This research was sponsored by UT Austin startup grant and the German Research Foundation (Deutsche For\-schungs\-gemein\-schaft, DFG) in the framework of Research Training Group \emph{GRK~2462: Processes in natural and technical Particle-Fluid-Systems} at Hamburg University of Technology (TUHH)\citep{PintPFS}.

\end{acknowledgements}

\bibliography{literature}  

\onecolumn

\end{document}